\documentclass[%
 aip,
 amsmath,amssymb,
 reprint,%
]{revtex4-2}

\usepackage{graphicx}
\usepackage{dcolumn}
\usepackage{bm}

\usepackage[utf8]{inputenc}
\usepackage[T1]{fontenc}
\usepackage{mathptmx}
\usepackage{etoolbox}
\usepackage[normalem]{ulem}

\makeatletter
\def\@email#1#2{%
 \endgroup
 \patchcmd{\titleblock@produce}
  {\frontmatter@RRAPformat}
  {\frontmatter@RRAPformat{\produce@RRAP{*#1\href{mailto:#2}{#2}}}\frontmatter@RRAPformat}
  {}{}
}%
\makeatother
\begin{document}

\preprint{AIP/123-QED}

\title{Cryogen-free scanning gate microscope for the characterization of Si/Si$_{0.7}$Ge$_{0.3}$ quantum devices at milli-Kelvin temperatures}
\author{Seong Woo Oh}\affiliation{Department of Physics,
  Princeton University, Princeton, New Jersey 08544, USA}
\author{Artem O. Denisov}\affiliation{Department of Physics,
  Princeton University, Princeton, New Jersey 08544, USA}
\author{Pengcheng Chen}\affiliation {Princeton Institute for the Science and Technology of Materials, Princeton University, Princeton, New Jersey 08544, USA}
\author{Jason R. Petta}\altaffiliation{Author to whom correspondence should be addressed; electronic mail: petta@princeton.edu.}\affiliation {Department of Physics, Princeton University, Princeton, New Jersey 08544, USA}

\date{\today}

\begin{abstract}
Silicon can be isotopically enriched, allowing for the fabrication of highly coherent semiconductor spin qubits. However, the conduction band of bulk Si exhibits a six-fold valley degeneracy, which may adversely impact the performance of silicon quantum devices. To date, the spatial characterization of valley states in Si remains limited. Moreover, techniques for probing valley states in functional electronic devices are needed. We describe here a cryogen-free scanning gate microscope for the characterization of Si/Si$_{0.7}$Ge$_{0.3}$ quantum devices at mK temperatures. The microscope is based on the Pan-walker design, with coarse positioning piezo stacks and a fine scanning piezo tube. A tungsten microscope tip is attached to a tuning fork for active control of the tip-to-sample distance. To reduce vibration noise from the pulse tube cooler, we utilize both active and passive vibration isolation mechanisms, and achieve a root-mean-square noise in $z$ of $\sim$ 2 nm. Our microscope is designed to characterize fully functioning Si/Si$_{0.7}$Ge$_{0.3}$ quantum devices. As a proof of concept, we use the microscope to manipulate the charge occupation of a Si quantum dot, opening up a range of possibilities for the exploration of quantum devices and materials. 
\end{abstract}

\maketitle

\section{Introduction}
Silicon spin qubits have rapidly evolved over the past decade and are now a legitimate contender in the race to build a scalable quantum computer~\cite{loss1998, Watson2018, Yang2018, Yoneda2018, Zajac2018, Mi2018}. The device fabrication process has matured, allowing for high yield and scale-up of modest one-dimensional Si quantum dot arrays~\cite{angus2007, doi:10.1063/1.4922249, ZajacScalable, Mills2018}. Recent experiments have demonstrated the suitability of Si/Si$_{0.7}$Ge$_{0.3}$ heterostructures as a platform for highly controllable Si spin qubits~\cite{Yoneda2018,andrews_quantifying_2019,Xue2018,Sigillito_2019}. Silicon's small intrinsic spin-orbit coupling and long spin coherence times, accompanied with well-established industrial fabrication process, have made Si/Si$_{0.7}$Ge$_{0.3}$ a promising platform for scalable quantum computing.

While Si has many favorable properties for the fabrication of semiconductor quantum devices, the bandstructure of bulk Si exhibits a six-fold ``valley degeneracy,'' which may introduce an uncontrolled orbital degree of freedom~\cite{schaffler_high-mobility_1997, Zwanenburg}. The tensile strain of the Si quantum well induced by the larger lattice constant of Ge in Si/Si$_{0.7}$Ge$_{0.3}$ heterostructures partially lifts the six-fold valley degeneracy by raising the in-plane valleys in energy relative to the $\pm z-$valleys~\cite{schaffler_high-mobility_1997}. It is the two lowest lying valleys that have posed one of the great challenges to silicon-based spin qubit technology~\cite{friesen_magnetic_2006,ftheory_2010}. Abrupt Si/Si$_{0.7}$Ge$_{0.3}$ interfaces can lift the two-fold degeneracy of the $z$-valleys, but in reality interfaces are not perfectly sharp \cite{Tariq,culcer_interface_2010,jiang_effects_2012,neyens_critical_2018}. Combined with atomic-scale disorder and step-edges, these effects lead to a large spread $25\rm\,  - 300\rm\, \mu eV$ in reported valley splittings \cite{borselli_measurement_2011,doi:10.1063/1.4922249,Mi_high_resolution,Borjans_Synthetic,ferdous_valley_2018,hollmann_large_2020,HRL_DAPS,PRXQuantum.2.020309}. Discovering a method
 to reliably engineer a large valley splitting would accelerate the development of a silicon-based quantum processor~\cite{Zwanenburg, penthorn, mcjunkin2021valley}.

\begin{figure*}[th] 
\includegraphics[width=2\columnwidth]{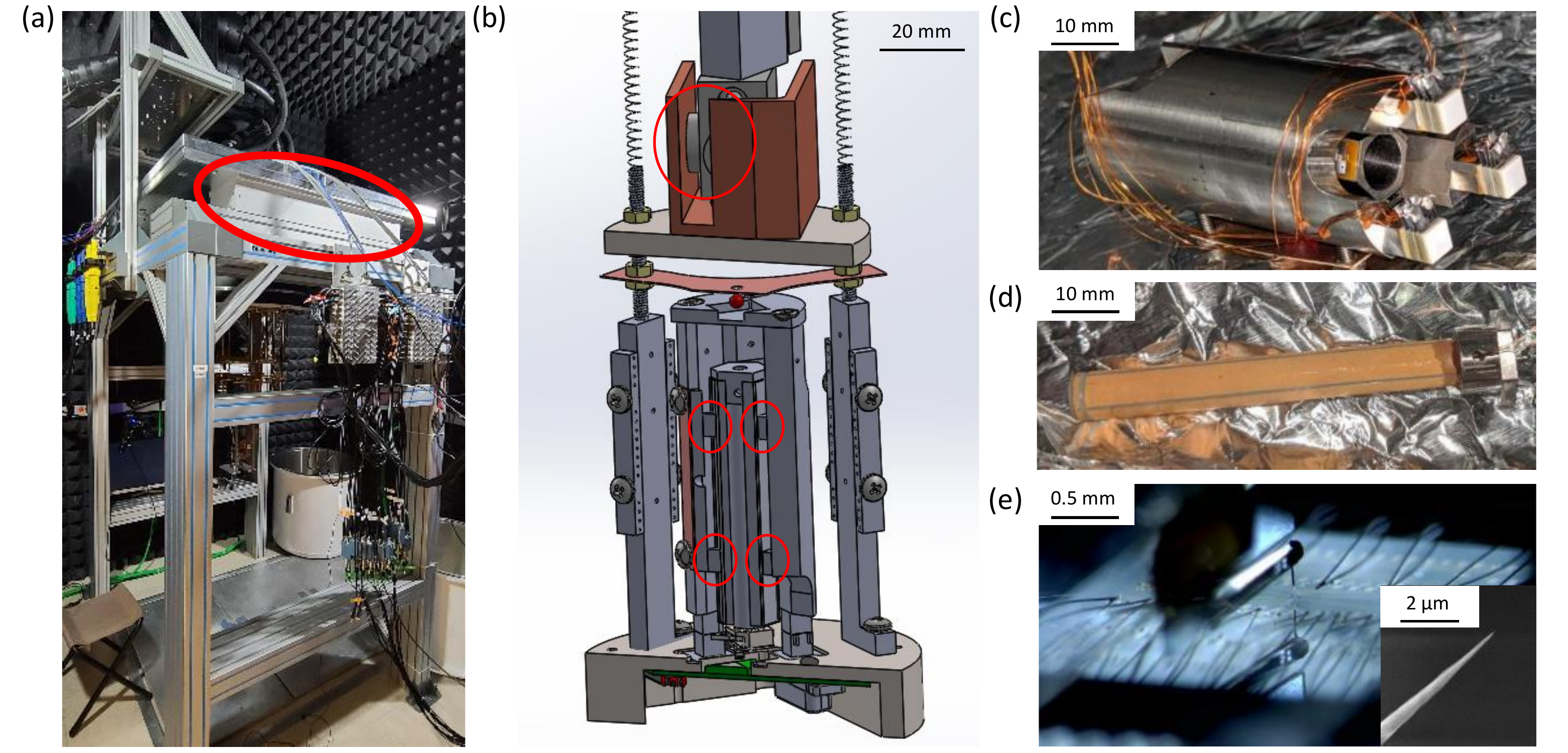}
\caption{Instrument overview. (a) The scanning gate microscope is installed beneath the mixing chamber plate of a BlueFors XLD dilution refrigerator. Active vibration dampers (circled in red) decouple the cryostat from the structural support stand. (b) Cross-section through the microscope, which consists of coarse $x$- and  $y$-positioners, a fine scanning tube, and a custom machined base plate that can accommodate a circuit board and quantum device. Sitting above the microscope is an open copper box with a block holding SmCo magnets for eddy current damping (top red circle). Circled in red below are four of the six coarse $z$-positioners. (c) Microscope body and coarse $xy$-positioners. (d) Fine scanning tube. (e) Optical image showing the tungsten scanning probe tip positioned above a Si/Si$_{0.7}$Ge$_{0.3}$ quantum device. The tungsten tip is attached to a Nanosurf S1.0 qPlus sensor and the sensor is glued to a rectangular chip that is pre-patterned with gold electrodes. A wire bonder is used to bond up the sensor lines. Inset: Scanning electron microscope (SEM) image of a tungsten tip that has been electrochemically etched for sensing purposes. 
}
	\label{fig1}
\end{figure*}

Looking beyond the materials challenges, progress has been impeded by a lack of a high throughput measurement of valley splitting. Valley splittings obtained from simple quantum Hall measurements are often much larger than those obtained from measurements on quantum dots~\cite{goswami_controllable_2007}, calling into question the utility of quantum Hall characterization. Quantum dot valley splittings are often extracted using magnetospectroscopy or are inferred from measurements of a spin-relaxation hot-spot that occurs when the Zeeman energy is comparable to the valley splitting~\cite{yang2013spin}. Recently, new valley splitting measurements have been developed that utilize single-shot singlet-triplet readout~\cite{HRL_singleshot}, pulsed detuning spectroscopy~\cite{HRL_DAPS}, or microwave spectroscopy in the circuit quantum electrodynamics architecture~\cite{Mi_high_resolution,PRXQuantum.2.020309}. All of these approaches require the time-consuming cycling through multiple devices to acquire meaningful statistics. Developing a scanning-probe measurement approach with spatial resolution and the ability to form a quantum dot at multiple locations on the same chip in a non-invasive manner would accelerate sample characterization.\cite{Tahan} 

In conjunction with improvements in quantum device fabrication, the range of scanning probe microscopy techniques has broadened considerably over the years. Conventional transport measurements, when combined with different scanning probe techniques, have opened up a range of capabilities including spatial mapping of branched electron flow~\cite{Topinka2001,Barnard2017} and the visualization of electron-hole puddles in graphene~\cite{Martin2008}. Applications of these capabilities in mesoscopic physics are abundant and include shot noise measurements of carrier charge~\cite{PhysRevB.100.104506} and the energy distribution function~\cite{PhysRevLett.79.3490, PhysRevB.102.085417}, as well as imaging of temperature gradients~\cite{Marguerite2019} and local magnetic fields~\cite{Uri2020,Ku2020,Nowack2013}. We envision the combination of scanning gate techniques with dispersive readout on-tip~\cite{Tahan} to allow spatial mapping of the valley splitting in Si/Si$_{0.7}$Ge$_{0.3}$ heterostructures. 

Here we describe the construction of a cryogen-free scanning gate microscope that is compatible with Si/Si$_{0.7}$Ge$_{0.3}$ quantum dot devices and operates at mK temperatures. The dimensions of the microscope parts, such as the scanning tube, sample holder printed circuit board, and microscope base, are chosen to be compatible with the dimensions of a BlueFors XLD dilution refrigerator. Where possible, we have heavily borrowed from existing STM/AFM designs~\cite{Pan_1999}. Several noise mitigating measures have been taken to reduce low frequency vibrations from the pulse tube. The article is divided into four technical sections. In Sec.\ II we give an overview of the microscope design parameters and mode of operation. The calibration of the microscope’s scanning piezo tube is described in Sec.\ III. In Sec.\ IV, we measure the noise in the microscope and make comparisons with existing cryogen-free systems. In Sec.\ V we demonstrate scanning gate microscopy of a Si/Si$_{0.7}$Ge$_{0.3}$ device, specifically using the microscope tip to control the charge occupation of a quantum dot.

\section{Instrumentation Design and Operation}

In this section several aspects of the microscope design are presented. Vibration isolation methods employed both inside and outside the fridge are detailed. We describe the Pan walker design principle, which enables the microscope to achieve coarse positioning. Then, we explain how the area of interest is located for imaging and device measurements. The scanning head design and assembly are outlined, as well as the printed circuit board (PCB) that enables transport measurements  while the scanning gate is rastered above the device. We also explain how the microscope tip approaches the sample in frequency modulated AFM (FM AFM) mode. 

\subsection{Vibration Isolation}

As opposed to high-end microscopy setups designed for experiments that require ultra-high vacuum and a high degree of noise suppression for subpicometer stability~\cite{doi:10.1063/1.3520482}, the setup described here is designed for versatile operation under rather harsh conditions. Our microscope is designed to be fully compatible with a BlueFors XLD cryogen-free dilution refrigerator. Several adjustments have been made to the standard BlueFors fridge installation configuration for the purpose of mitigating the mechanical noise from the pulse tube. The remote option is used to decouple the pulse tube motors from the pulse tubes (Cryomech PT-415), and has been shown to suppress to a large degree the vibrations that would otherwise transfer to the mixing chamber plate of the fridge~\cite{Pelliccione_2013}. In addition, heavy aluminum slabs on the top of the fridge frame provide additional stability. An AVI-400/LP active damper [see Fig.~\ref{fig1}(a)] is employed between the cryostat support frame and the aluminum slabs on which the fridge rests as a cost-effective alternative to more intricate vibration damping systems employed in other setups. The BlueFors cryostat is housed in a shielded room with walls lined with extruded melamine foam to reduce ambient acoustic noise.

Additional vibration mitigation measures are taken below the mixing chamber plate of the dilution refrigerator. Linear BeCu suspension springs mechanically decouple the microscope from the mixing chamber plate of the cryostat. An eddy current damper provides additional stability and is constructed by rigidly attaching a rectangular titanium piece to the bottom of the mixing chamber plate. The lower end of the Ti piece has samarium-cobalt magnets attached to each side [top red circle in Fig.~\ref{fig1}(b)]. These magnets face the inner walls of a Cu box that is mounted to the top of the microscope to create the magnetic damping effect. An annealed copper braid is used to thermally link the microscope to the mixing chamber plate. 

\subsection{Coarse Positioning}

To a large extent, the microscope is based on the well-known Pan design~\cite{Pan_1999} with a slight modification. The overall actuation mechanism for the $x$-, $y$-, and $z$-directions is as follows. In the original Pan-walker scheme, to move the load (microscope) in the $x$-direction, for example, the piezo controller sends high-voltage pulses sequentially to $x$-piezo stacks in order to move the piezos one at a time while other stacks hold the load in place. After the sequence of pulses is complete to displace the resting position of the load, the piezo controller slowly decreases the applied voltages to zero. However, this configuration requires a pair of wires for each piezo stack. In order to simplify the wiring scheme, most modern microscopes typically assign only one pair of wires to each set of $x$-, $y$-, and $z$-piezo stacks. This approach leads to degraded coarse positioning performance, especially at low temperatures. Therefore, instead of assigning the same actuation voltage to all six $z$-piezo stacks, we group them into two sets. The top set of the $z$-piezo stacks, two of which are shown in the middle red circles of Fig.\ 1(b), and the bottom set of the $z$-piezo stacks, two of which are shown in the bottom red circles of Fig.\ 1(b), receive voltage pulses with an adjustable time delay for trouble free motion.   

Materials science-focused microscopes often forgo lateral positioning for the sake of mechanical stability. However, for our application to quantum devices,  the ability to traverse the surface of a sample is important since our chips are millimeters in scale, and the device area is on the order of several microns in scale. Therefore, in spite of the partial loss of mechanical stability, we adopt a walking mechanism based on $xy$-piezo stacks~\cite{Hug_1999}. As shown in Fig.~\ref{fig1}(c), three $xy$-piezo stacks, glued to the rectangular legs of the microscope's main body, walk across sapphire plates on the titanium base. With the Nanonis SPM controller's piezo motor controller (PMD4 version D) one can program an arbitrary time delay between the two sets of piezo motors. As explained in the previous paragraph, this time delay has been shown to be effective at preventing the $z$-piezo positioners from freezing at low temperatures. Both the $xy$- and $z$-piezo positioners are held in contact with sapphire plates using thin BeCu sheets as springs [BeCu sheet sitting above the red ruby ball in Fig.~\ref{fig1}(b)]. The microscope slides across the sapphire plates for coarse positioning with respect to the area of interest on the device. Additionally, we note that if the pressure on the $z$-piezo motors is excessive, the $z$-direction actuation tends to freeze at low temperatures. For optimal operation, the spring tension is adjusted at room temperature such that the ratio between the number of steps we need to take up and down to walk the same $z$ distance is roughly 1.38 (here, a step means one voltage pulse sent to the $z$-piezos). 

\subsection{Scanning}

Fine scanning of the sample is achieved using a piezo tube [see Fig.~\ref{fig1}(d)], which is housed inside the cylindrical body shown in Fig.~\ref{fig1}(c). Generally, for mechanical stability and compactness of the microscope, the typical length of a piezo tube is around 25 mm or less. The total length extension linearly increases with the length of the scanning tube as $\Delta L = \frac{d_{31}VL}{t}$, where $d_{31}$ is the vertical extension coefficient, $V$ is the voltage, $L$ is the natural length of the tube, and $t$ is the tube wall thickness. We choose a longer scanning tube (50 mm) in order to safely deal with $\sim20-150~\mathrm{nm}$ multilayer Al gate structure of Si/Si$_{0.7}$Ge$_{0.3}$ devices~\cite{doi:10.1063/1.4922249}. In addition, the scan area available without activating coarse positioners is linearly dependent on the length of the piezo tube as $\Delta x = \Delta y =\frac{0.9d_{31}VL^2}{d_m t}$, where $d_m$ is the average of the inner and outer diameters of the scanning tube, and $t$ is the thickness of the scanning tube~\cite{EBL}. We chose 50 mm, 6.35 mm, and 0.5 mm for the scanning tube length, outer diameter, and thickness, respectively.  

\subsection{Scanning Head (qPlus sensor and Tip)}

A sensor in our typical configuration is shown in Fig.~\ref{fig1}(e). We use the S1.0 Nanosurf qPlus sensor as our mechanical resonator~\cite{Giessibl}. As opposed to a quartz tuning fork with two oscillating prongs, a qPlus sensor has only one freely oscillating prong. qPlus sensors exhibit cantilever-like dynamics and a lower effective elastic constant $k_{\rm eff}$. However tuning fork sensors can have a much higher quality factor $Q$ and produce at least twice the piezoelectric voltage of a qPlus sensor for a given tip oscillation amplitude~\cite{Cast}. Our decision to use the Nanosurf qPlus sensor is mostly due to the convenience of having a dedicated electrical lead on the resonator for the scanning probe tip.  

To prepare a scanning head, we first glue a qPlus sensor onto a square quartz chip with non-conductive epoxy. The quartz chip has been pre-patterned with Ti/Au electrodes and the far ends of the Ti/Au electrodes are connected to enameled copper wires. Two of the wires are used to read out the qPlus resonator oscillation signal, and the remaining wire is used to apply voltage to the probe tip. Once the epoxy is cured in an oven at 375 $^{\circ}$F for 20 min., we use a wirebonder to make electrical connections between the electrodes on the quartz chip and the qPlus sensor. We then glue a 25 $\mu$m diameter tungsten wire with conductive epoxy to the free prong of the qPlus sensor, while ensuring that the wire is in electrical contact with the qPlus sensor's probe electrical lead. The tungsten tip is electrochemically etched following a commonly used recipe~\cite{oliva}.

Although the declared resonance frequency of the tuning fork is 32,768~Hz, with a metallic tip glued to the free-standing prong of the resonator, the actual resonance frequency is below 30~kHz. The base of the quartz chip, which has the qPlus sensor glued on it, is next glued to a ceramic plate. The ceramic plate has BeCu springs attached to it. The two enameled wires on the quartz chip for the qPlus oscillation voltage readout are glued to the BeCu springs. Then, the packaged sensor is plugged into the sensor housing located on the end of the scanning tube as shown in Fig.~\ref{fig1}(d). The BeCu springs offer mechanical stability inside the sensor housing and also make electrical contact with a pair of stainless steel wires in order to provide the means for the qPlus sensor signal readout. The electrical wires are threaded out of the sensor housing to a cryogenic amplifier that is mounted on the cold plate (around 750~mK) of the pulse tube dilution refrigerator.

\subsection{Sample Holder}

A custom-made PCB is used as a sample holder. The PCB has one Glenair connector with 25 pins, 24 of which are used to electrically bias gate electrodes and apply signals to ohmic contacts on the device. Standard wedgebonding is used to make connections between bond pads on the chip and the conductive traces on the PCB. In addition to the usual design considerations for Si/Si$_{0.7}$Ge$_{0.3}$ device circuit boards, there are two extra dimensional requirements. First, the width of the PCB has to be narrow enough to fit under the base of the microscope without interfering with the bore of the vector magnet. Second, the device sits on a pedestal relative to the other surfaces of the PCB, allowing for optical access and rough alignment of the tip to the active area of the device. The result is a wedding cake PCB design with the motherboard (1 mm thick) and pedestal (1.7 mm thick) consisting of two copper layers each. 

\subsection{Tip-Sample Approach}

The PCB sample holder is designed to accomodate chips of size $6 \times 6$ mm. Each chip typically has four devices on it, with the electrically active $\sim$ 5 $\times$ 5 $\mu$m area located in the center of each device. Given the absence of a viewport in the BlueFors XLD fridge, we use a long working distance microscope to achieve rough alignment of the scanning probe to the active area of the device. A long working distance microscope is attached to the frame of the BlueFors fridge for this purpose. With a qPlus sensor packaged with a sharp metallic tip and inserted in the sensor housing, and a device mounted underneath the base of the microscope, we perform a rough alignment between the tip and the device by actuating the coarse $xyz$-piezo positioners. The result of such a procedure is shown in Fig.~\ref{fig1}(e). When the rough alignment is complete, the bottom loading cryostat is closed and pumped down. Once the vacuum level reaches a reasonable level (below 1 $\times 10^{-2}$~mBar), the qPlus sensor can start approaching the device at room temperature. 

We approach the device in FM AFM mode. In this mode of operation, the phase locked loop (PLL) of the microscope controller tracks the resonance frequency of the qPlus sensor. When the tip gets close to the device, the mechanical resonance frequency shifts due to the force interaction with the surface. Once the resonance frequency shift reaches the setpoint, the touchdown is complete. Then, a few AFM topographic images are taken at the room temperature for the purpose of locating the center of the device. After each scan, the $xy$-coarse piezo positioners are actuated to position the tip of the sensor closer to the center of the device. After the iterative positioning process, the device is found, and the cryostat is ready to be cooled down to base temperature ($\sim$20 mK). We optimize the PLL and proportional-integral (PI) settings of the AFM controller once we reach base temperature. However, we compensate for the lateral thermal drift of the microscope during the cooldown to mK temperatures by repeating a topographic scan and adjusting the lateral position of the microscope at 4~K.

\begin{figure}[t] 
\includegraphics[width=\columnwidth]{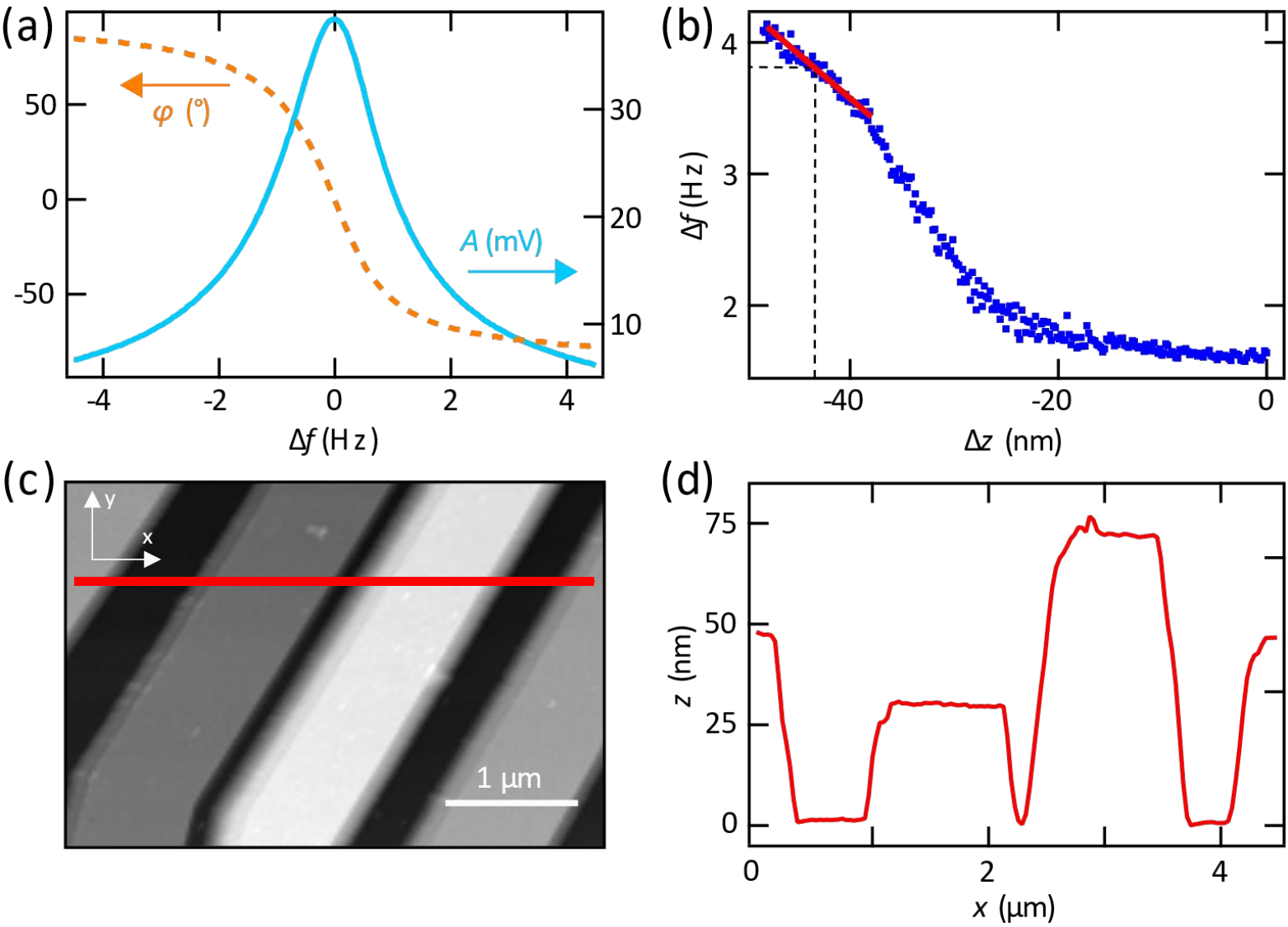}
\caption{Operation of the microscope in AFM mode. (a) Typical amplitude $A$ and phase $\phi$ of the qPlus sensor signal at 20 mK as a function of frequency offset $\Delta f$. The center frequency is 27.499 kHz and the quality factor $Q$ $\sim$ 17,600. (b) $z$-spectroscopy is used to determine the frequency shift-to-distance conversion factor ($-6.433\times10^{7} $ Hz/m) around the setpoint frequency $\Delta f$ = 3.625 Hz. (c) AFM image of a three-layer aluminum gate stack fabricated on a silicon surface. (d) A line-cut through the data in (c), showing the heights of the different gates.
}
	\label{fig2}
\end{figure}

\section{Instrumentation Calibration}

In order to scan a sample and characterize the noise in the tip-to-sample separation (detailed description in the next section), one must first choose a mechanical resonance frequency to track. We first identify the mechanical resonances of the tuning fork by measuring the voltage and phase across the tuning fork as a function of drive frequency $f$ as shown in Fig.~\ref{fig2}(a). A tuning fork has multiple resonances with varying quality $Q$ factors. For a scanning probe system with a pulse tube cooler, it is important to optimize the PI loop parameters to deal with external mechanical noise, especially when sample features are on the order of tens of nm or more. High $Q$ resonances tend to react more drastically to external noise and height variations during scans, which lead to a higher chance of damaging the tip. Therefore, we went about identifying optimal PI loop parameters. In our findings, the optimal P range is on the order of nm/Hz, and the optimal I range tends to be around hundreds of nm/Hz/sec. These settings allow for reliable imaging with a high $Q\sim15,000$ resonance. 

After finding the main resonance of the tuning fork, we scan the gates of the device in order to acquire the room temperature-to-mK scaling factor for the scanning piezo tube. The expansion coefficient of the scanning tube piezo varies as a function of temperature, such that the scanning tube piezo requires a higher voltage for the same amount of expansion at low temperatures~\cite{EBL}. Figure \ref{fig2}(c) shows an image of four metallic gates. Knowing the heights of the gates, we determined that the room temperature-to-mK scaling factor is roughly 7.46 in the $z$-direction. During imaging, the microscope control software adjusts the voltage applied to the scanning tube piezo in order to track the height variation of the surface. The applied voltage is then converted to meters based on the distance to input voltage conversion factor. The scaling factor that we found above is used to adjust this conversion factor appropriately. From this point on, the measured $z$-position is now scaled according to the temperature dependence of the scanning piezo tube.

To characterize the vibrations in the tip-sample distance in the absence of a feedback loop we measure the shift in the tuning fork resonance frequency as a function of tip-sample $z$-separation, also known as $z$-spectroscopy. The steps taken to acquire the plot shown in Fig.\ 2(b) are as follows: i.) Choose a setpoint frequency (3.68 Hz, in this specific case) where the scanning piezo tube stabilizes. ii.) Note how much the scanning tube piezo had to expand in order to reach the setpoint frequency and choose the $z$-sweep range appropriately, so that the frequency shift includes the setpoint while the tip does not crash into the sample surface for the entire sweep range. iii.) Perform multiple sweeps over the selected $z$-range while measuring the shift of the tuning fork resonance frequency and signal average. Figure~\ref{fig2}(b) shows the raw data acquired from $z$-spectroscopy. We extract the frequency-to-distance conversion factor in Hz/nm from the linear fit around the frequency setpoint [see dashed lines in Fig. \ref{fig2}(b)]. This conversion factor is required to convert the frequency noise power spectral density (units of Hz/$\sqrt{\text{Hz}}$) to tip-to-sample noise power spectral density in (units of nm/$\sqrt{\text{Hz}}$).

\begin{figure}[t] 
\includegraphics[width=\columnwidth]{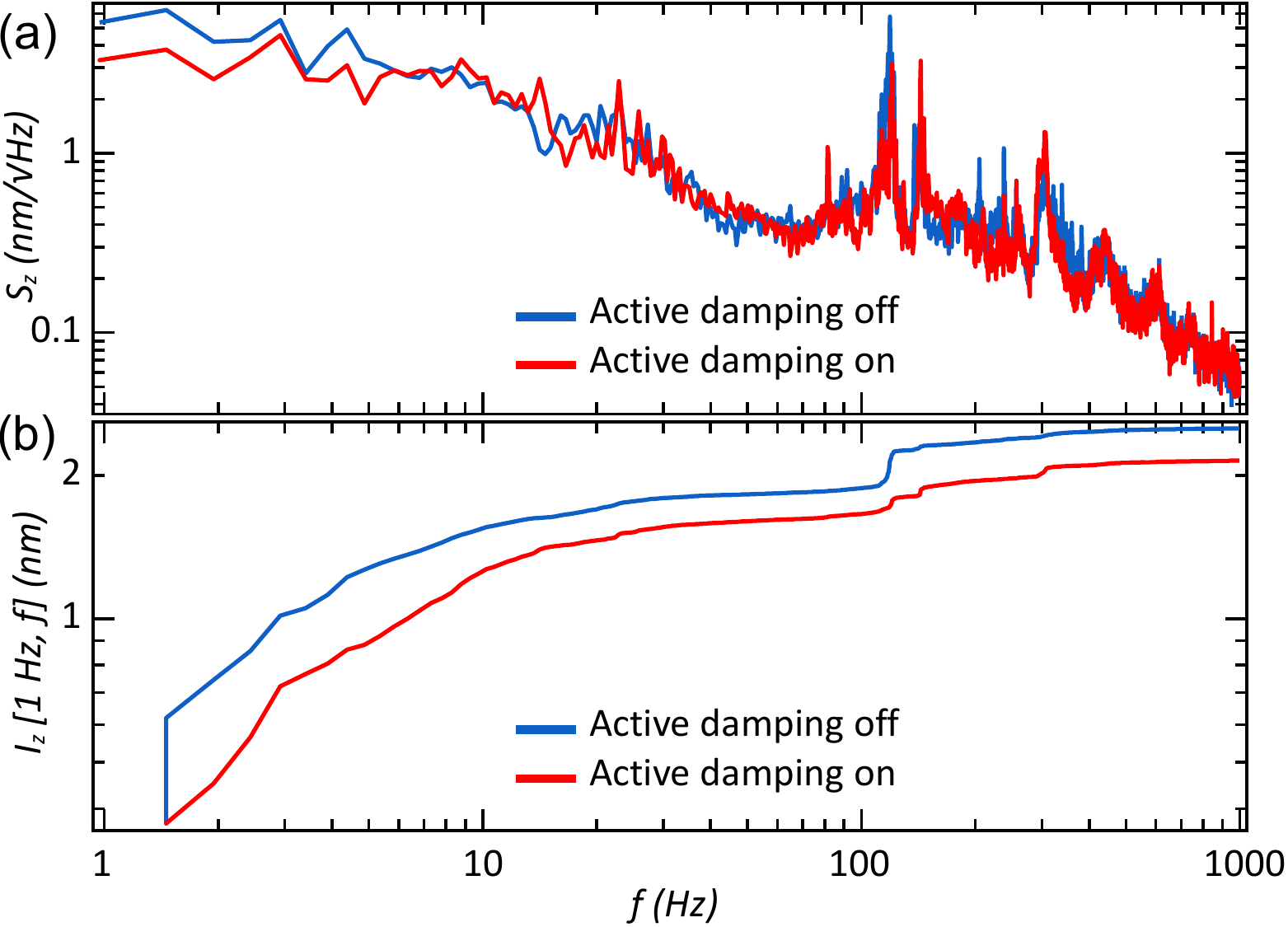}
\caption{Noise analysis. (a) $z$-position noise power power spectral density $S_\mathrm{Z}$ acquired with the active damping system (AVI-400/LP) turned off and on. (b) Integrated noise $I_\mathrm{Z}$ over a frequency range of 1 to 1000~Hz with active damping turned off and on.}
	\label{fig3}
\end{figure}

\section{Performance Evaluation}

We quantitatively analyze the microscope's noise characteristics by measuring the power spectral density of the tuning fork resonance frequency. Since the frequency-to-distance conversion factor from the $z$-spectroscopy performed in the previous section is acquired from the linear fit around the frequency of 3.68~Hz, the frequency power spectral density is measured at the setpoint of 3.68~Hz in order to be able to accurately convert the frequency fluctuation to the distance fluctuation equivalent. The resulting tip-to-surface power spectral density is shown in Fig.~\ref{fig3}(a). In order to test the effectiveness of the active vibration damper, we measure the tip-to-surface power spectral density with the damper on and off. For easier comparison, we integrate the noise spectra to acquire the root-mean-square (rms) vibration noise, defined as \begin{math} I_{\mathrm{Z}}[1, f] = \sqrt{\int_{1}^{f} [S(f^{'})]^2 df^{'} }\end{math}. The integrated noise plotted in Fig.~\ref{fig3}(b) shows that the main contributions to the noise occur at around 120, 140 and 300~Hz. Overall, the damping system reduces the noise by $16\%$. We find the noise level of our system to be comparable to other setups~\cite{Pelliccione_2013,Kalisky,Geaney2019}, which have reported 2.1 nm or more for the tip-to-sample vibration.

\section{Scanning Gate Microscopy}
\begin{figure}[t] 
\includegraphics[width=\columnwidth]{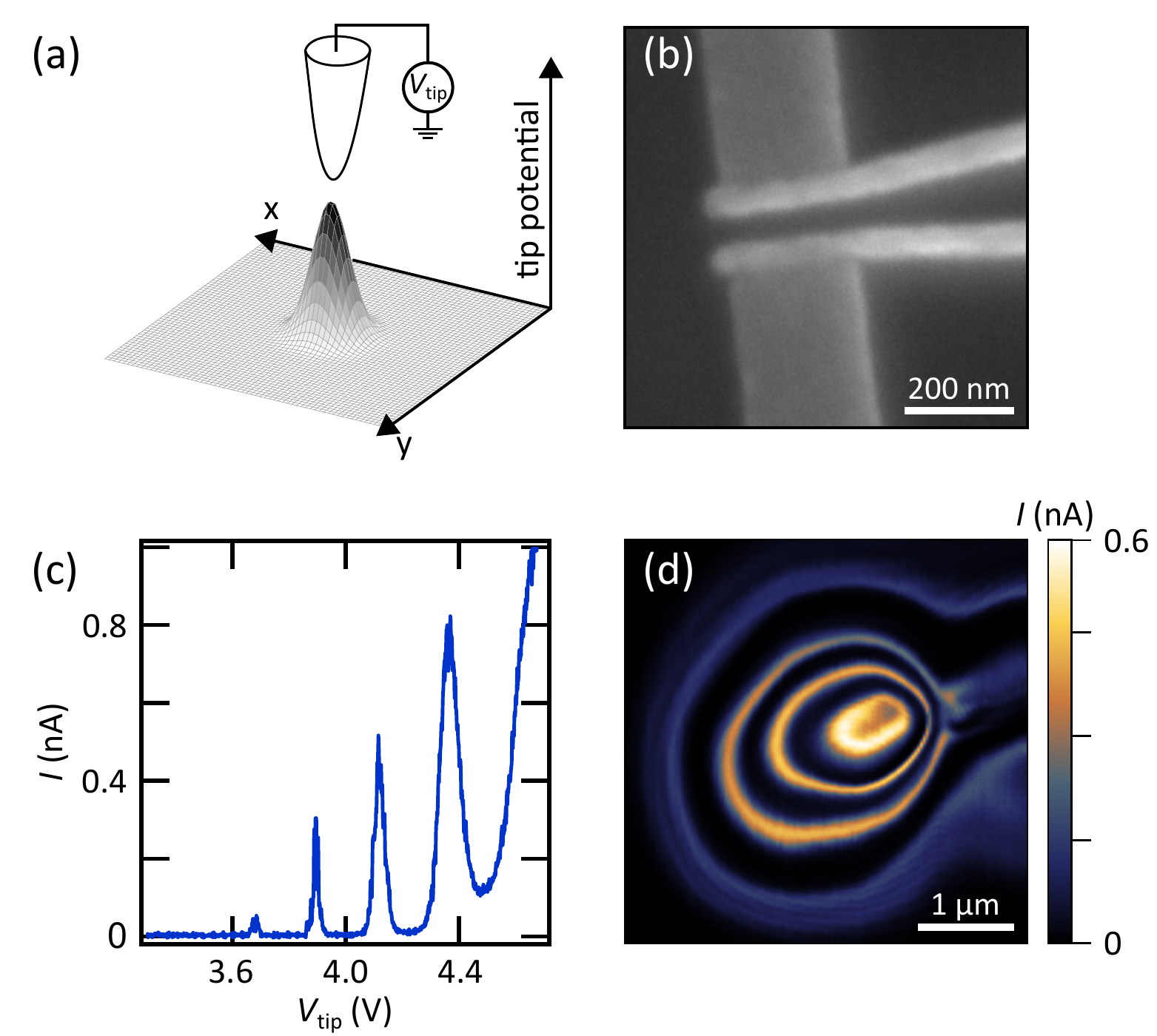}
\caption{Demonstration of scanning gate microscopy on a Si/Si$_{0.7}$Ge$_{0.3}$ quantum device. (a) Schematic of the local potential induced by the tip with a negative voltage bias. (b) SEM image of the double-layer device. (c) Coulumb peaks as a function of the tip voltage measured at constant gate voltages. The tip is parked roughly 100~nm above the surface of the wafer near the center of the device. (d) Scanning gate microsocopy. Measured currrent as a function of the tip position with the tip voltage $V_{\mathrm{tip}}= 0.5 ~V$. Notice that device tuning is now different from (c). No filtering or data smoothing is used. All data are acquired with a 20~mK lattice temperature.}
	\label{fig4}
\end{figure}

We now demonstrate full operation of the microscope by imaging an electrically active Si/Si$_{0.7}$Ge$_{0.3}$ quantum device at mK temperatures. A scanning gate microscopy experiment is performed by scanning the tungsten tip, which is dc-biased at a potential $V_{\mathrm{tip}}$ relative to the substrate, above the active area of the device~\cite{Woodside_2002, PhysRevLett.93.216801, PhysRevLett.91.246803}. The sample consists of a 11 nm thick Si quantum well that is buried by a 20 nm thick layer of Si$_{0.7}$Ge$_{0.3}$ and capped by 2 nm of Si~\cite{xiao_mobility}. Figure~\ref{fig4}(a) illustrates the potential induced by the negatively biased tip in the buried Si quantum well. The tip induced potential can be changed by adjusting $V_{\mathrm{tip}}$ or the tip-device distance. Changing the $xy$-position of the tip around a gate-defined quantum dot (QD) affects the QD confinement potential and shifts the QD energy levels~\cite{RevModPhys.79.1217}. Scanning-gate microscopy is the mapping of the current or conductance through the QD as a function the tip position.

The SEM image of the device made on the Si/Si$_{0.7}$Ge$_{0.3}$ heterostructure~\cite{doi:10.1063/1.4922249} is shown in the inset of Fig.~\ref{fig4}(b). By applying a voltage $V_{\mathrm{G}}$ to two Al barrier gates we can form a QD between the pair of accumulation gates. We intentionally removed a plunger gate from the device to demonstrate that we can use the tip instead to change the electron occupation of the QD. To that end, we fix the tip approximately 100~nm above the wafer surface near the center of the device and turn off the dither excitation and the feedback loop. In Fig.~\ref{fig4}(c) single electron transport is evident from the Coulumb peaks in the measured dc current $I$ which is plotted as a function of $V_{\mathrm{tip}}$. The other gate voltages are held fixed during this measurement and we apply a source - drain bias $V_{\rm sd}~\sim150~\mu$V across the device.

The scanning gate image of the QD is shown in Fig.~\ref{fig4}(d). We adjusted the barrier gate voltage $V_{\mathrm{G}}$ to reach few electron occupancy of the QD with the tip close and $V_{\mathrm{tip}}=0.5~V$. The image is acquired as the tip is scanned in the plane around 100~nm above the device with zero dither excitation and a constant tip voltage. As expected \cite{Woodside_2002, PhysRevLett.93.216801, Fallahi2005}, the concentric rings observed in the scanning gate image correspond to Coulomb blockade peaks in the measured current. Each ring corresponds to an equipotential line. Moving in the radial direction towards the center of the dot, the crossing of each ring corresponds to the addition of an electron to the dot.
 
\section{CONCLUSIONS}
We have designed, constructed, and evaluated the performance of a cryogen-free scanning gate microscope capable of operation at mK temperatures. Active vibration isolation is achieved at room temperature and pulse tube noise is further supressed at mK temperatures using passive vibration isolation. The scanning probe consists of a dc-biased tungsten tip that is attached to a commercial qPlus tuning fork. Imaging over large areas is made possible using coarse $xyz$-piezo positioners. Fine scanning in the active area of the device is achieved using a scanning tube in a Pan microscope configuration \cite{Pan_1999}. Measurements of the noise power spectral density indicate a rms noise in the $z$ tip-sample separation of $\sim$ 2 nm, comparable to other cryogen free microscopes. Our system is designed to be compatible with fully functional quantum devices. As an example, we use the constructed microscope to perform a scanning gate experiment on a quantum dot defined in a Si/Si$_{0.7}$Ge$_{0.3}$ heterostructure. Future efforts will be directed at measuring the spatial variation of the valley splitting in Si/Si$_{0.7}$Ge$_{0.3}$ quantum dots.

\section{ACKNOWLEDGEMENTS}
We thank Arthur Barnard, David Goldhaber-Gordon, Matthew Pelliccione, Rick Silver, and Joe Stroscio for helpful discussions, and Sonia Zhang for technical contributions to device fabrication. We would also like to thank Dr.\ Jarosik and Dr.\ Lengel for AFM related insights and Nanonis support. Construction of the microscope was made possible by the Gordon and Betty Moore Foundation’s EPiQS Initiative through grant GBMF4535, with partial support from Army Research Office grant W911NF-15-1-0149. Devices were fabricated in the Princeton University Quantum Device Nanofabrication Laboratory, which is managed by the department of physics.

\bibliography{References_Seong_RSI_2021_v5}{}

\begin{thebibliography}{60}%
\makeatletter
\providecommand \@ifxundefined [1]{%
 \@ifx{#1\undefined}
}%
\providecommand \@ifnum [1]{%
 \ifnum #1\expandafter \@firstoftwo
 \else \expandafter \@secondoftwo
 \fi
}%
\providecommand \@ifx [1]{%
 \ifx #1\expandafter \@firstoftwo
 \else \expandafter \@secondoftwo
 \fi
}%
\providecommand \natexlab [1]{#1}%
\providecommand \enquote  [1]{``#1''}%
\providecommand \bibnamefont  [1]{#1}%
\providecommand \bibfnamefont [1]{#1}%
\providecommand \citenamefont [1]{#1}%
\providecommand \href@noop [0]{\@secondoftwo}%
\providecommand \href [0]{\begingroup \@sanitize@url \@href}%
\providecommand \@href[1]{\@@startlink{#1}\@@href}%
\providecommand \@@href[1]{\endgroup#1\@@endlink}%
\providecommand \@sanitize@url [0]{\catcode `\\12\catcode `\$12\catcode
  `\&12\catcode `\#12\catcode `\^12\catcode `\_12\catcode `\%12\relax}%
\providecommand \@@startlink[1]{}%
\providecommand \@@endlink[0]{}%
\providecommand \url  [0]{\begingroup\@sanitize@url \@url }%
\providecommand \@url [1]{\endgroup\@href {#1}{\urlprefix }}%
\providecommand \urlprefix  [0]{URL }%
\providecommand \Eprint [0]{\href }%
\providecommand \doibase [0]{https://doi.org/}%
\providecommand \selectlanguage [0]{\@gobble}%
\providecommand \bibinfo  [0]{\@secondoftwo}%
\providecommand \bibfield  [0]{\@secondoftwo}%
\providecommand \translation [1]{[#1]}%
\providecommand \BibitemOpen [0]{}%
\providecommand \bibitemStop [0]{}%
\providecommand \bibitemNoStop [0]{.\EOS\space}%
\providecommand \EOS [0]{\spacefactor3000\relax}%
\providecommand \BibitemShut  [1]{\csname bibitem#1\endcsname}%
\let\auto@bib@innerbib\@empty
\bibitem [{\citenamefont {Loss}\ and\ \citenamefont
  {DiVincenzo}(1998)}]{loss1998}%
  \BibitemOpen
  \bibfield  {author} {\bibinfo {author} {\bibfnamefont {D.}~\bibnamefont
  {Loss}}\ and\ \bibinfo {author} {\bibfnamefont {D.~P.}\ \bibnamefont
  {DiVincenzo}},\ }\bibfield  {title} {\enquote {\bibinfo {title} {Quantum
  computation with quantum dots},}\ }\href
  {https://doi.org/10.1103/PhysRevA.57.120} {\bibfield  {journal} {\bibinfo
  {journal} {Phys. Rev. A}\ }\textbf {\bibinfo {volume} {57}},\ \bibinfo
  {pages} {120} (\bibinfo {year} {1998})}\BibitemShut {NoStop}%
\bibitem [{\citenamefont {Watson}\ \emph {et~al.}(2018)\citenamefont {Watson},
  \citenamefont {Philips}, \citenamefont {Kawakami}, \citenamefont {Ward},
  \citenamefont {Scarlino}, \citenamefont {Veldhorst}, \citenamefont {Savage},
  \citenamefont {Lagally}, \citenamefont {Friesen}, \citenamefont
  {Coppersmith}, \citenamefont {Eriksson},\ and\ \citenamefont
  {Vandersypen}}]{Watson2018}%
  \BibitemOpen
  \bibfield  {author} {\bibinfo {author} {\bibfnamefont {T.~F.}\ \bibnamefont
  {Watson}}, \bibinfo {author} {\bibfnamefont {S.~G.~J.}\ \bibnamefont
  {Philips}}, \bibinfo {author} {\bibfnamefont {E.}~\bibnamefont {Kawakami}},
  \bibinfo {author} {\bibfnamefont {D.~R.}\ \bibnamefont {Ward}}, \bibinfo
  {author} {\bibfnamefont {P.}~\bibnamefont {Scarlino}}, \bibinfo {author}
  {\bibfnamefont {M.}~\bibnamefont {Veldhorst}}, \bibinfo {author}
  {\bibfnamefont {D.~E.}\ \bibnamefont {Savage}}, \bibinfo {author}
  {\bibfnamefont {M.~G.}\ \bibnamefont {Lagally}}, \bibinfo {author}
  {\bibfnamefont {M.}~\bibnamefont {Friesen}}, \bibinfo {author} {\bibfnamefont
  {S.~N.}\ \bibnamefont {Coppersmith}}, \bibinfo {author} {\bibfnamefont
  {M.~A.}\ \bibnamefont {Eriksson}},\ and\ \bibinfo {author} {\bibfnamefont
  {L.~M.~K.}\ \bibnamefont {Vandersypen}},\ }\bibfield  {title} {\enquote
  {\bibinfo {title} {A programmable two-qubit quantum processor in silicon},}\
  }\href {https://doi.org/10.1038/nature25766} {\bibfield  {journal} {\bibinfo
  {journal} {Nature (London)}\ }\textbf {\bibinfo {volume} {555}},\ \bibinfo
  {pages} {633} (\bibinfo {year} {2018})}\BibitemShut {NoStop}%
\bibitem [{\citenamefont {{Yang}}\ \emph {et~al.}(2019)\citenamefont {{Yang}},
  \citenamefont {{Chan}}, \citenamefont {{Harper}}, \citenamefont {{Huang}},
  \citenamefont {{Evans}}, \citenamefont {{Hwang}}, \citenamefont {{Hensen}},
  \citenamefont {{Laucht}}, \citenamefont {{Tanttu}}, \citenamefont {{Hudson}},
  \citenamefont {{Flammia}}, \citenamefont {{Itoh}}, \citenamefont {{Morello}},
  \citenamefont {{Bartlett}},\ and\ \citenamefont {{Dzurak}}}]{Yang2018}%
  \BibitemOpen
  \bibfield  {author} {\bibinfo {author} {\bibfnamefont {C.~H.}\ \bibnamefont
  {{Yang}}}, \bibinfo {author} {\bibfnamefont {K.~W.}\ \bibnamefont {{Chan}}},
  \bibinfo {author} {\bibfnamefont {R.}~\bibnamefont {{Harper}}}, \bibinfo
  {author} {\bibfnamefont {W.}~\bibnamefont {{Huang}}}, \bibinfo {author}
  {\bibfnamefont {T.}~\bibnamefont {{Evans}}}, \bibinfo {author} {\bibfnamefont
  {J.~C.~C.}\ \bibnamefont {{Hwang}}}, \bibinfo {author} {\bibfnamefont
  {B.}~\bibnamefont {{Hensen}}}, \bibinfo {author} {\bibfnamefont
  {A.}~\bibnamefont {{Laucht}}}, \bibinfo {author} {\bibfnamefont
  {T.}~\bibnamefont {{Tanttu}}}, \bibinfo {author} {\bibfnamefont {F.~E.}\
  \bibnamefont {{Hudson}}}, \bibinfo {author} {\bibfnamefont {S.~T.}\
  \bibnamefont {{Flammia}}}, \bibinfo {author} {\bibfnamefont {K.~M.}\
  \bibnamefont {{Itoh}}}, \bibinfo {author} {\bibfnamefont {A.}~\bibnamefont
  {{Morello}}}, \bibinfo {author} {\bibfnamefont {S.~D.}\ \bibnamefont
  {{Bartlett}}},\ and\ \bibinfo {author} {\bibfnamefont {A.~S.}\ \bibnamefont
  {{Dzurak}}},\ }\bibfield  {title} {\enquote {\bibinfo {title} {{Silicon qubit
  fidelities approaching incoherent noise limits via pulse optimisation}},}\
  }\href@noop {} {\bibfield  {journal} {\bibinfo  {journal} {Nat. Electron.}\
  }\textbf {\bibinfo {volume} {2}},\ \bibinfo {pages} {151} (\bibinfo {year}
  {2019})}\BibitemShut {NoStop}%
\bibitem [{\citenamefont {Yoneda}\ \emph {et~al.}(2018)\citenamefont {Yoneda},
  \citenamefont {Takeda}, \citenamefont {Otsuka}, \citenamefont {Nakajima},
  \citenamefont {Delbecq}, \citenamefont {Allison}, \citenamefont {Honda},
  \citenamefont {Kodera}, \citenamefont {Oda}, \citenamefont {Hoshi},
  \citenamefont {Usami}, \citenamefont {Itoh},\ and\ \citenamefont
  {Tarucha}}]{Yoneda2018}%
  \BibitemOpen
  \bibfield  {author} {\bibinfo {author} {\bibfnamefont {J.}~\bibnamefont
  {Yoneda}}, \bibinfo {author} {\bibfnamefont {K.}~\bibnamefont {Takeda}},
  \bibinfo {author} {\bibfnamefont {T.}~\bibnamefont {Otsuka}}, \bibinfo
  {author} {\bibfnamefont {T.}~\bibnamefont {Nakajima}}, \bibinfo {author}
  {\bibfnamefont {M.~R.}\ \bibnamefont {Delbecq}}, \bibinfo {author}
  {\bibfnamefont {G.}~\bibnamefont {Allison}}, \bibinfo {author} {\bibfnamefont
  {T.}~\bibnamefont {Honda}}, \bibinfo {author} {\bibfnamefont
  {T.}~\bibnamefont {Kodera}}, \bibinfo {author} {\bibfnamefont
  {S.}~\bibnamefont {Oda}}, \bibinfo {author} {\bibfnamefont {Y.}~\bibnamefont
  {Hoshi}}, \bibinfo {author} {\bibfnamefont {N.}~\bibnamefont {Usami}},
  \bibinfo {author} {\bibfnamefont {K.~M.}\ \bibnamefont {Itoh}},\ and\
  \bibinfo {author} {\bibfnamefont {S.}~\bibnamefont {Tarucha}},\ }\bibfield
  {title} {\enquote {\bibinfo {title} {A quantum-dot spin qubit with coherence
  limited by charge noise and fidelity higher than 99.9\%},}\ }\href
  {https://doi.org/10.1038/s41565-017-0014-x} {\bibfield  {journal} {\bibinfo
  {journal} {Nat. Nanotechnol.}\ }\textbf {\bibinfo {volume} {13}},\ \bibinfo
  {pages} {102} (\bibinfo {year} {2018})}\BibitemShut {NoStop}%
\bibitem [{\citenamefont {Zajac}\ \emph {et~al.}(2018)\citenamefont {Zajac},
  \citenamefont {Sigillito}, \citenamefont {Russ}, \citenamefont {Borjans},
  \citenamefont {Taylor}, \citenamefont {Burkard},\ and\ \citenamefont
  {Petta}}]{Zajac2018}%
  \BibitemOpen
  \bibfield  {author} {\bibinfo {author} {\bibfnamefont {D.~M.}\ \bibnamefont
  {Zajac}}, \bibinfo {author} {\bibfnamefont {A.~J.}\ \bibnamefont
  {Sigillito}}, \bibinfo {author} {\bibfnamefont {M.}~\bibnamefont {Russ}},
  \bibinfo {author} {\bibfnamefont {F.}~\bibnamefont {Borjans}}, \bibinfo
  {author} {\bibfnamefont {J.~M.}\ \bibnamefont {Taylor}}, \bibinfo {author}
  {\bibfnamefont {G.}~\bibnamefont {Burkard}},\ and\ \bibinfo {author}
  {\bibfnamefont {J.~R.}\ \bibnamefont {Petta}},\ }\bibfield  {title} {\enquote
  {\bibinfo {title} {Resonantly driven {CNOT} gate for electron spins},}\
  }\href {https://doi.org/10.1126/science.aao5965} {\bibfield  {journal}
  {\bibinfo  {journal} {Science}\ }\textbf {\bibinfo {volume} {359}},\ \bibinfo
  {pages} {439} (\bibinfo {year} {2018})}\BibitemShut {NoStop}%
\bibitem [{\citenamefont {Mi}\ \emph {et~al.}(2018)\citenamefont {Mi},
  \citenamefont {Benito}, \citenamefont {Putz}, \citenamefont {Zajac},
  \citenamefont {Taylor}, \citenamefont {Burkard},\ and\ \citenamefont
  {Petta}}]{Mi2018}%
  \BibitemOpen
  \bibfield  {author} {\bibinfo {author} {\bibfnamefont {X.}~\bibnamefont
  {Mi}}, \bibinfo {author} {\bibfnamefont {M.}~\bibnamefont {Benito}}, \bibinfo
  {author} {\bibfnamefont {S.}~\bibnamefont {Putz}}, \bibinfo {author}
  {\bibfnamefont {D.~M.}\ \bibnamefont {Zajac}}, \bibinfo {author}
  {\bibfnamefont {J.~M.}\ \bibnamefont {Taylor}}, \bibinfo {author}
  {\bibfnamefont {G.}~\bibnamefont {Burkard}},\ and\ \bibinfo {author}
  {\bibfnamefont {J.~R.}\ \bibnamefont {Petta}},\ }\bibfield  {title} {\enquote
  {\bibinfo {title} {{A coherent spin–photon interface in silicon}},}\ }\href
  {http://dx.doi.org/10.1038/nature25769} {\bibfield  {journal} {\bibinfo
  {journal} {Nature (London)}\ }\textbf {\bibinfo {volume} {555}},\ \bibinfo
  {pages} {599} (\bibinfo {year} {2018})}\BibitemShut {NoStop}%
\bibitem [{\citenamefont {Angus}\ \emph {et~al.}(2007)\citenamefont {Angus},
  \citenamefont {Ferguson}, \citenamefont {Dzurak},\ and\ \citenamefont
  {Clark}}]{angus2007}%
  \BibitemOpen
  \bibfield  {author} {\bibinfo {author} {\bibfnamefont {S.~J.}\ \bibnamefont
  {Angus}}, \bibinfo {author} {\bibfnamefont {A.~J.}\ \bibnamefont {Ferguson}},
  \bibinfo {author} {\bibfnamefont {A.~S.}\ \bibnamefont {Dzurak}},\ and\
  \bibinfo {author} {\bibfnamefont {R.~G.}\ \bibnamefont {Clark}},\ }\bibfield
  {title} {\enquote {\bibinfo {title} {Gate-defined quantum dots in intrinsic
  silicon},}\ }\href@noop {} {\bibfield  {journal} {\bibinfo  {journal} {Nano
  Lett.}\ }\textbf {\bibinfo {volume} {7}},\ \bibinfo {pages} {2051} (\bibinfo
  {year} {2007})}\BibitemShut {NoStop}%
\bibitem [{\citenamefont {Zajac}\ \emph {et~al.}(2015)\citenamefont {Zajac},
  \citenamefont {Hazard}, \citenamefont {Mi}, \citenamefont {Wang},\ and\
  \citenamefont {Petta}}]{doi:10.1063/1.4922249}%
  \BibitemOpen
  \bibfield  {author} {\bibinfo {author} {\bibfnamefont {D.~M.}\ \bibnamefont
  {Zajac}}, \bibinfo {author} {\bibfnamefont {T.~M.}\ \bibnamefont {Hazard}},
  \bibinfo {author} {\bibfnamefont {X.}~\bibnamefont {Mi}}, \bibinfo {author}
  {\bibfnamefont {K.}~\bibnamefont {Wang}},\ and\ \bibinfo {author}
  {\bibfnamefont {J.~R.}\ \bibnamefont {Petta}},\ }\bibfield  {title} {\enquote
  {\bibinfo {title} {{A reconfigurable gate architecture for Si/SiGe quantum
  dots}},}\ }\href@noop {} {\bibfield  {journal} {\bibinfo  {journal} {Appl.
  Phys. Lett.}\ }\textbf {\bibinfo {volume} {106}},\ \bibinfo {pages} {223507}
  (\bibinfo {year} {2015})}\BibitemShut {NoStop}%
\bibitem [{\citenamefont {Zajac}\ \emph {et~al.}(2016)\citenamefont {Zajac},
  \citenamefont {Hazard}, \citenamefont {Mi}, \citenamefont {Nielsen},\ and\
  \citenamefont {Petta}}]{ZajacScalable}%
  \BibitemOpen
  \bibfield  {author} {\bibinfo {author} {\bibfnamefont {D.~M.}\ \bibnamefont
  {Zajac}}, \bibinfo {author} {\bibfnamefont {T.~M.}\ \bibnamefont {Hazard}},
  \bibinfo {author} {\bibfnamefont {X.}~\bibnamefont {Mi}}, \bibinfo {author}
  {\bibfnamefont {E.}~\bibnamefont {Nielsen}},\ and\ \bibinfo {author}
  {\bibfnamefont {J.~R.}\ \bibnamefont {Petta}},\ }\bibfield  {title} {\enquote
  {\bibinfo {title} {Scalable gate architecture for a one-dimensional array of
  semiconductor spin qubits},}\ }\href
  {https://doi.org/10.1103/PhysRevApplied.6.054013} {\bibfield  {journal}
  {\bibinfo  {journal} {Phys. Rev. Appl.}\ }\textbf {\bibinfo {volume} {6}},\
  \bibinfo {pages} {054013} (\bibinfo {year} {2016})}\BibitemShut {NoStop}%
\bibitem [{\citenamefont {{Mills}}\ \emph {et~al.}(2019)\citenamefont
  {{Mills}}, \citenamefont {{Zajac}}, \citenamefont {{Gullans}}, \citenamefont
  {{Schupp}}, \citenamefont {{Hazard}},\ and\ \citenamefont
  {{Petta}}}]{Mills2018}%
  \BibitemOpen
  \bibfield  {author} {\bibinfo {author} {\bibfnamefont {A.~R.}\ \bibnamefont
  {{Mills}}}, \bibinfo {author} {\bibfnamefont {D.~M.}\ \bibnamefont
  {{Zajac}}}, \bibinfo {author} {\bibfnamefont {M.~J.}\ \bibnamefont
  {{Gullans}}}, \bibinfo {author} {\bibfnamefont {F.~J.}\ \bibnamefont
  {{Schupp}}}, \bibinfo {author} {\bibfnamefont {T.~M.}\ \bibnamefont
  {{Hazard}}},\ and\ \bibinfo {author} {\bibfnamefont {J.~R.}\ \bibnamefont
  {{Petta}}},\ }\bibfield  {title} {\enquote {\bibinfo {title} {{Shuttling a
  single charge across a one-dimensional array of silicon quantum dots}},}\
  }\href@noop {} {\bibfield  {journal} {\bibinfo  {journal} {Nat. Commun.}\
  }\textbf {\bibinfo {volume} {10}},\ \bibinfo {pages} {1063} (\bibinfo {year}
  {2019})}\BibitemShut {NoStop}%
\bibitem [{\citenamefont {Andrews}\ \emph {et~al.}(2019)\citenamefont
  {Andrews}, \citenamefont {Jones}, \citenamefont {Reed}, \citenamefont
  {Jones}, \citenamefont {Ha}, \citenamefont {Jura}, \citenamefont {Kerckhoff},
  \citenamefont {Levendorf}, \citenamefont {Meenehan}, \citenamefont {Merkel},
  \citenamefont {Smith}, \citenamefont {Sun}, \citenamefont {Weinstein},
  \citenamefont {Rakher}, \citenamefont {Ladd},\ and\ \citenamefont
  {Borselli}}]{andrews_quantifying_2019}%
  \BibitemOpen
  \bibfield  {author} {\bibinfo {author} {\bibfnamefont {R.~W.}\ \bibnamefont
  {Andrews}}, \bibinfo {author} {\bibfnamefont {C.}~\bibnamefont {Jones}},
  \bibinfo {author} {\bibfnamefont {M.~D.}\ \bibnamefont {Reed}}, \bibinfo
  {author} {\bibfnamefont {A.~M.}\ \bibnamefont {Jones}}, \bibinfo {author}
  {\bibfnamefont {S.~D.}\ \bibnamefont {Ha}}, \bibinfo {author} {\bibfnamefont
  {M.~P.}\ \bibnamefont {Jura}}, \bibinfo {author} {\bibfnamefont
  {J.}~\bibnamefont {Kerckhoff}}, \bibinfo {author} {\bibfnamefont
  {M.}~\bibnamefont {Levendorf}}, \bibinfo {author} {\bibfnamefont
  {S.}~\bibnamefont {Meenehan}}, \bibinfo {author} {\bibfnamefont {S.~T.}\
  \bibnamefont {Merkel}}, \bibinfo {author} {\bibfnamefont {A.}~\bibnamefont
  {Smith}}, \bibinfo {author} {\bibfnamefont {B.}~\bibnamefont {Sun}}, \bibinfo
  {author} {\bibfnamefont {A.~J.}\ \bibnamefont {Weinstein}}, \bibinfo {author}
  {\bibfnamefont {M.~T.}\ \bibnamefont {Rakher}}, \bibinfo {author}
  {\bibfnamefont {T.~D.}\ \bibnamefont {Ladd}},\ and\ \bibinfo {author}
  {\bibfnamefont {M.~G.}\ \bibnamefont {Borselli}},\ }\bibfield  {title}
  {\enquote {\bibinfo {title} {Quantifying error and leakage in an encoded
  {Si}/{SiGe} triple-dot qubit},}\ }\href
  {https://doi.org/10.1038/s41565-019-0500-4} {\bibfield  {journal} {\bibinfo
  {journal} {Nat. Nanotechnol.}\ }\textbf {\bibinfo {volume} {14}},\ \bibinfo
  {pages} {747} (\bibinfo {year} {2019})}\BibitemShut {NoStop}%
\bibitem [{\citenamefont {{Xue}}\ \emph {et~al.}(2019)\citenamefont {{Xue}},
  \citenamefont {{Watson}}, \citenamefont {{Helsen}}, \citenamefont {{Ward}},
  \citenamefont {{Savage}}, \citenamefont {{Lagally}}, \citenamefont
  {{Coppersmith}}, \citenamefont {{Eriksson}}, \citenamefont {{Wehner}},\ and\
  \citenamefont {{Vandersypen}}}]{Xue2018}%
  \BibitemOpen
  \bibfield  {author} {\bibinfo {author} {\bibfnamefont {X.}~\bibnamefont
  {{Xue}}}, \bibinfo {author} {\bibfnamefont {T.~F.}\ \bibnamefont {{Watson}}},
  \bibinfo {author} {\bibfnamefont {J.}~\bibnamefont {{Helsen}}}, \bibinfo
  {author} {\bibfnamefont {D.~R.}\ \bibnamefont {{Ward}}}, \bibinfo {author}
  {\bibfnamefont {D.~E.}\ \bibnamefont {{Savage}}}, \bibinfo {author}
  {\bibfnamefont {M.~G.}\ \bibnamefont {{Lagally}}}, \bibinfo {author}
  {\bibfnamefont {S.~N.}\ \bibnamefont {{Coppersmith}}}, \bibinfo {author}
  {\bibfnamefont {M.~A.}\ \bibnamefont {{Eriksson}}}, \bibinfo {author}
  {\bibfnamefont {S.}~\bibnamefont {{Wehner}}},\ and\ \bibinfo {author}
  {\bibfnamefont {L.~M.~K.}\ \bibnamefont {{Vandersypen}}},\ }\bibfield
  {title} {\enquote {\bibinfo {title} {{Benchmarking Gate Fidelities in a
  Si/SiGe Two-Qubit Device}},}\ }\href@noop {} {\bibfield  {journal} {\bibinfo
  {journal} {Phys. Rev. X}\ }\textbf {\bibinfo {volume} {9}},\ \bibinfo {pages}
  {021011} (\bibinfo {year} {2019})}\BibitemShut {NoStop}%
\bibitem [{\citenamefont {Sigillito}\ \emph {et~al.}(2019)\citenamefont
  {Sigillito}, \citenamefont {Loy}, \citenamefont {Zajac}, \citenamefont
  {Gullans}, \citenamefont {Edge},\ and\ \citenamefont
  {Petta}}]{Sigillito_2019}%
  \BibitemOpen
  \bibfield  {author} {\bibinfo {author} {\bibfnamefont {A.~J.}\ \bibnamefont
  {Sigillito}}, \bibinfo {author} {\bibfnamefont {J.~C.}\ \bibnamefont {Loy}},
  \bibinfo {author} {\bibfnamefont {D.~M.}\ \bibnamefont {Zajac}}, \bibinfo
  {author} {\bibfnamefont {M.~J.}\ \bibnamefont {Gullans}}, \bibinfo {author}
  {\bibfnamefont {L.~F.}\ \bibnamefont {Edge}},\ and\ \bibinfo {author}
  {\bibfnamefont {J.~R.}\ \bibnamefont {Petta}},\ }\bibfield  {title} {\enquote
  {\bibinfo {title} {Site-{S}elective {Q}uantum {C}ontrol in an {I}sotopically
  {E}nriched $\mathrm{{}^{28}Si}$/$\mathrm{Si}_{0.7}\mathrm{Ge}_{0.3}$
  {Q}uadruple {Q}uantum {D}ot},}\ }\href
  {https://doi.org/10.1103/PhysRevApplied.11.061006} {\bibfield  {journal}
  {\bibinfo  {journal} {Phys. Rev. Appl.}\ }\textbf {\bibinfo {volume} {11}},\
  \bibinfo {pages} {061006} (\bibinfo {year} {2019})}\BibitemShut {NoStop}%
\bibitem [{\citenamefont {Schäffler}(1997)}]{schaffler_high-mobility_1997}%
  \BibitemOpen
  \bibfield  {author} {\bibinfo {author} {\bibfnamefont {F.}~\bibnamefont
  {Schäffler}},\ }\bibfield  {title} {\enquote {\bibinfo {title}
  {High-mobility {Si} and {Ge} structures},}\ }\href
  {https://doi.org/10.1088/0268-1242/12/12/001} {\bibfield  {journal} {\bibinfo
   {journal} {Semicond. Sci. Technol.}\ }\textbf {\bibinfo {volume} {12}},\
  \bibinfo {pages} {1515} (\bibinfo {year} {1997})}\BibitemShut {NoStop}%
\bibitem [{\citenamefont {Zwanenburg}\ \emph {et~al.}(2013)\citenamefont
  {Zwanenburg}, \citenamefont {Dzurak}, \citenamefont {Morello}, \citenamefont
  {Simmons}, \citenamefont {Hollenberg}, \citenamefont {Klimeck}, \citenamefont
  {Rogge}, \citenamefont {Coppersmith},\ and\ \citenamefont
  {Eriksson}}]{Zwanenburg}%
  \BibitemOpen
  \bibfield  {author} {\bibinfo {author} {\bibfnamefont {F.~A.}\ \bibnamefont
  {Zwanenburg}}, \bibinfo {author} {\bibfnamefont {A.~S.}\ \bibnamefont
  {Dzurak}}, \bibinfo {author} {\bibfnamefont {A.}~\bibnamefont {Morello}},
  \bibinfo {author} {\bibfnamefont {M.~Y.}\ \bibnamefont {Simmons}}, \bibinfo
  {author} {\bibfnamefont {L.~C.~L.}\ \bibnamefont {Hollenberg}}, \bibinfo
  {author} {\bibfnamefont {G.}~\bibnamefont {Klimeck}}, \bibinfo {author}
  {\bibfnamefont {S.}~\bibnamefont {Rogge}}, \bibinfo {author} {\bibfnamefont
  {S.~N.}\ \bibnamefont {Coppersmith}},\ and\ \bibinfo {author} {\bibfnamefont
  {M.~A.}\ \bibnamefont {Eriksson}},\ }\bibfield  {title} {\enquote {\bibinfo
  {title} {Silicon quantum electronics},}\ }\href
  {https://doi.org/10.1103/RevModPhys.85.961} {\bibfield  {journal} {\bibinfo
  {journal} {Rev. Mod. Phys.}\ }\textbf {\bibinfo {volume} {85}},\ \bibinfo
  {pages} {961} (\bibinfo {year} {2013})}\BibitemShut {NoStop}%
\bibitem [{\citenamefont {Friesen}, \citenamefont {Eriksson},\ and\
  \citenamefont {Coppersmith}(2006)}]{friesen_magnetic_2006}%
  \BibitemOpen
  \bibfield  {author} {\bibinfo {author} {\bibfnamefont {M.}~\bibnamefont
  {Friesen}}, \bibinfo {author} {\bibfnamefont {M.~A.}\ \bibnamefont
  {Eriksson}},\ and\ \bibinfo {author} {\bibfnamefont {S.~N.}\ \bibnamefont
  {Coppersmith}},\ }\bibfield  {title} {\enquote {\bibinfo {title} {Magnetic
  field dependence of valley splitting in realistic {Si}/{SiGe} quantum
  wells},}\ }\href {https://doi.org/10.1063/1.2387975} {\bibfield  {journal}
  {\bibinfo  {journal} {Appl. Phys. Lett.}\ }\textbf {\bibinfo {volume} {89}},\
  \bibinfo {pages} {202106} (\bibinfo {year} {2006})}\BibitemShut {NoStop}%
\bibitem [{\citenamefont {Friesen}\ and\ \citenamefont
  {Coppersmith}(2010)}]{ftheory_2010}%
  \BibitemOpen
  \bibfield  {author} {\bibinfo {author} {\bibfnamefont {M.}~\bibnamefont
  {Friesen}}\ and\ \bibinfo {author} {\bibfnamefont {S.~N.}\ \bibnamefont
  {Coppersmith}},\ }\bibfield  {title} {\enquote {\bibinfo {title} {Theory of
  valley-orbit coupling in a {Si}/{SiGe} quantum dot},}\ }\href
  {https://doi.org/10.1103/PhysRevB.81.115324} {\bibfield  {journal} {\bibinfo
  {journal} {Phys. Rev. B}\ }\textbf {\bibinfo {volume} {81}},\ \bibinfo
  {pages} {115324} (\bibinfo {year} {2010})}\BibitemShut {NoStop}%
\bibitem [{\citenamefont {Tariq}\ and\ \citenamefont {Hu}(2019)}]{Tariq}%
  \BibitemOpen
  \bibfield  {author} {\bibinfo {author} {\bibfnamefont {B.}~\bibnamefont
  {Tariq}}\ and\ \bibinfo {author} {\bibfnamefont {X.}~\bibnamefont {Hu}},\
  }\bibfield  {title} {\enquote {\bibinfo {title} {Effects of interface steps
  on the valley-orbit coupling in a {Si}/{SiGe} quantum dot},}\ }\href
  {https://doi.org/10.1103/PhysRevB.100.125309} {\bibfield  {journal} {\bibinfo
   {journal} {Phys. Rev. B}\ }\textbf {\bibinfo {volume} {100}},\ \bibinfo
  {pages} {125309} (\bibinfo {year} {2019})}\BibitemShut {NoStop}%
\bibitem [{\citenamefont {Culcer}, \citenamefont {Hu},\ and\ \citenamefont
  {Das~Sarma}(2010)}]{culcer_interface_2010}%
  \BibitemOpen
  \bibfield  {author} {\bibinfo {author} {\bibfnamefont {D.}~\bibnamefont
  {Culcer}}, \bibinfo {author} {\bibfnamefont {X.}~\bibnamefont {Hu}},\ and\
  \bibinfo {author} {\bibfnamefont {S.}~\bibnamefont {Das~Sarma}},\ }\bibfield
  {title} {\enquote {\bibinfo {title} {Interface roughness, valley-orbit
  coupling, and valley manipulation in quantum dots},}\ }\href
  {https://doi.org/10.1103/PhysRevB.82.205315} {\bibfield  {journal} {\bibinfo
  {journal} {Phys. Rev. B}\ }\textbf {\bibinfo {volume} {82}},\ \bibinfo
  {pages} {205315} (\bibinfo {year} {2010})}\BibitemShut {NoStop}%
\bibitem [{\citenamefont {Jiang}\ \emph {et~al.}(2012)\citenamefont {Jiang},
  \citenamefont {Kharche}, \citenamefont {Boykin},\ and\ \citenamefont
  {Klimeck}}]{jiang_effects_2012}%
  \BibitemOpen
  \bibfield  {author} {\bibinfo {author} {\bibfnamefont {Z.}~\bibnamefont
  {Jiang}}, \bibinfo {author} {\bibfnamefont {N.}~\bibnamefont {Kharche}},
  \bibinfo {author} {\bibfnamefont {T.}~\bibnamefont {Boykin}},\ and\ \bibinfo
  {author} {\bibfnamefont {G.}~\bibnamefont {Klimeck}},\ }\bibfield  {title}
  {\enquote {\bibinfo {title} {Effects of interface disorder on valley
  splitting in {SiGe}/{Si}/{SiGe} quantum wells},}\ }\href
  {https://doi.org/10.1063/1.3692174} {\bibfield  {journal} {\bibinfo
  {journal} {Appl. Phys. Lett.}\ }\textbf {\bibinfo {volume} {100}},\ \bibinfo
  {pages} {103502} (\bibinfo {year} {2012})}\BibitemShut {NoStop}%
\bibitem [{\citenamefont {Neyens}\ \emph {et~al.}(2018)\citenamefont {Neyens},
  \citenamefont {Foote}, \citenamefont {Thorgrimsson}, \citenamefont {Knapp},
  \citenamefont {McJunkin}, \citenamefont {Vandersypen}, \citenamefont {Amin},
  \citenamefont {Thomas}, \citenamefont {Clarke}, \citenamefont {Savage},
  \citenamefont {Lagally}, \citenamefont {Friesen}, \citenamefont
  {Coppersmith},\ and\ \citenamefont {Eriksson}}]{neyens_critical_2018}%
  \BibitemOpen
  \bibfield  {author} {\bibinfo {author} {\bibfnamefont {S.~F.}\ \bibnamefont
  {Neyens}}, \bibinfo {author} {\bibfnamefont {R.~H.}\ \bibnamefont {Foote}},
  \bibinfo {author} {\bibfnamefont {B.}~\bibnamefont {Thorgrimsson}}, \bibinfo
  {author} {\bibfnamefont {T.~J.}\ \bibnamefont {Knapp}}, \bibinfo {author}
  {\bibfnamefont {T.}~\bibnamefont {McJunkin}}, \bibinfo {author}
  {\bibfnamefont {L.~M.~K.}\ \bibnamefont {Vandersypen}}, \bibinfo {author}
  {\bibfnamefont {P.}~\bibnamefont {Amin}}, \bibinfo {author} {\bibfnamefont
  {N.~K.}\ \bibnamefont {Thomas}}, \bibinfo {author} {\bibfnamefont {J.~S.}\
  \bibnamefont {Clarke}}, \bibinfo {author} {\bibfnamefont {D.~E.}\
  \bibnamefont {Savage}}, \bibinfo {author} {\bibfnamefont {M.~G.}\
  \bibnamefont {Lagally}}, \bibinfo {author} {\bibfnamefont {M.}~\bibnamefont
  {Friesen}}, \bibinfo {author} {\bibfnamefont {S.~N.}\ \bibnamefont
  {Coppersmith}},\ and\ \bibinfo {author} {\bibfnamefont {M.~A.}\ \bibnamefont
  {Eriksson}},\ }\bibfield  {title} {\enquote {\bibinfo {title} {The critical
  role of substrate disorder in valley splitting in {Si} quantum wells},}\
  }\href {https://doi.org/10.1063/1.5033447} {\bibfield  {journal} {\bibinfo
  {journal} {Appl. Phys. Lett.}\ }\textbf {\bibinfo {volume} {112}},\ \bibinfo
  {pages} {243107} (\bibinfo {year} {2018})}\BibitemShut {NoStop}%
\bibitem [{\citenamefont {Borselli}\ \emph {et~al.}(2011)\citenamefont
  {Borselli}, \citenamefont {Ross}, \citenamefont {Kiselev}, \citenamefont
  {Croke}, \citenamefont {Holabird}, \citenamefont {Deelman}, \citenamefont
  {Warren}, \citenamefont {Alvarado-Rodriguez}, \citenamefont {Milosavljevic},
  \citenamefont {Ku}, \citenamefont {Wong}, \citenamefont {Schmitz},
  \citenamefont {Sokolich}, \citenamefont {Gyure},\ and\ \citenamefont
  {Hunter}}]{borselli_measurement_2011}%
  \BibitemOpen
  \bibfield  {author} {\bibinfo {author} {\bibfnamefont {M.~G.}\ \bibnamefont
  {Borselli}}, \bibinfo {author} {\bibfnamefont {R.~S.}\ \bibnamefont {Ross}},
  \bibinfo {author} {\bibfnamefont {A.~A.}\ \bibnamefont {Kiselev}}, \bibinfo
  {author} {\bibfnamefont {E.~T.}\ \bibnamefont {Croke}}, \bibinfo {author}
  {\bibfnamefont {K.~S.}\ \bibnamefont {Holabird}}, \bibinfo {author}
  {\bibfnamefont {P.~W.}\ \bibnamefont {Deelman}}, \bibinfo {author}
  {\bibfnamefont {L.~D.}\ \bibnamefont {Warren}}, \bibinfo {author}
  {\bibfnamefont {I.}~\bibnamefont {Alvarado-Rodriguez}}, \bibinfo {author}
  {\bibfnamefont {I.}~\bibnamefont {Milosavljevic}}, \bibinfo {author}
  {\bibfnamefont {F.~C.}\ \bibnamefont {Ku}}, \bibinfo {author} {\bibfnamefont
  {W.~S.}\ \bibnamefont {Wong}}, \bibinfo {author} {\bibfnamefont {A.~E.}\
  \bibnamefont {Schmitz}}, \bibinfo {author} {\bibfnamefont {M.}~\bibnamefont
  {Sokolich}}, \bibinfo {author} {\bibfnamefont {M.~F.}\ \bibnamefont
  {Gyure}},\ and\ \bibinfo {author} {\bibfnamefont {A.~T.}\ \bibnamefont
  {Hunter}},\ }\bibfield  {title} {\enquote {\bibinfo {title} {Measurement of
  valley splitting in high-symmetry {Si}/{SiGe} quantum dots},}\ }\href
  {https://doi.org/10.1063/1.3569717} {\bibfield  {journal} {\bibinfo
  {journal} {Appl. Phys. Lett.}\ }\textbf {\bibinfo {volume} {98}},\ \bibinfo
  {pages} {123118} (\bibinfo {year} {2011})}\BibitemShut {NoStop}%
\bibitem [{\citenamefont {Mi}\ \emph {et~al.}(2017)\citenamefont {Mi},
  \citenamefont {P\'eterfalvi}, \citenamefont {Burkard},\ and\ \citenamefont
  {Petta}}]{Mi_high_resolution}%
  \BibitemOpen
  \bibfield  {author} {\bibinfo {author} {\bibfnamefont {X.}~\bibnamefont
  {Mi}}, \bibinfo {author} {\bibfnamefont {C.~G.}\ \bibnamefont
  {P\'eterfalvi}}, \bibinfo {author} {\bibfnamefont {G.}~\bibnamefont
  {Burkard}},\ and\ \bibinfo {author} {\bibfnamefont {J.~R.}\ \bibnamefont
  {Petta}},\ }\bibfield  {title} {\enquote {\bibinfo {title} {{High-Resolution
  Valley Spectroscopy of Si Quantum Dots}},}\ }\href
  {https://doi.org/10.1103/PhysRevLett.119.176803} {\bibfield  {journal}
  {\bibinfo  {journal} {Phys. Rev. Lett.}\ }\textbf {\bibinfo {volume} {119}},\
  \bibinfo {pages} {176803} (\bibinfo {year} {2017})}\BibitemShut {NoStop}%
\bibitem [{\citenamefont {Borjans}\ \emph {et~al.}(2019)\citenamefont
  {Borjans}, \citenamefont {Zajac}, \citenamefont {Hazard},\ and\ \citenamefont
  {Petta}}]{Borjans_Synthetic}%
  \BibitemOpen
  \bibfield  {author} {\bibinfo {author} {\bibfnamefont {F.}~\bibnamefont
  {Borjans}}, \bibinfo {author} {\bibfnamefont {D.}~\bibnamefont {Zajac}},
  \bibinfo {author} {\bibfnamefont {T.}~\bibnamefont {Hazard}},\ and\ \bibinfo
  {author} {\bibfnamefont {J.}~\bibnamefont {Petta}},\ }\bibfield  {title}
  {\enquote {\bibinfo {title} {Single-spin relaxation in a synthetic spin-orbit
  field},}\ }\href {https://doi.org/10.1103/PhysRevApplied.11.044063}
  {\bibfield  {journal} {\bibinfo  {journal} {Phys. Rev. Applied}\ }\textbf
  {\bibinfo {volume} {11}},\ \bibinfo {pages} {044063} (\bibinfo {year}
  {2019})}\BibitemShut {NoStop}%
\bibitem [{\citenamefont {Ferdous}\ \emph {et~al.}(2018)\citenamefont
  {Ferdous}, \citenamefont {Kawakami}, \citenamefont {Scarlino}, \citenamefont
  {Nowak}, \citenamefont {Ward}, \citenamefont {Savage}, \citenamefont
  {Lagally}, \citenamefont {Coppersmith}, \citenamefont {Friesen},
  \citenamefont {Eriksson}, \citenamefont {Vandersypen},\ and\ \citenamefont
  {Rahman}}]{ferdous_valley_2018}%
  \BibitemOpen
  \bibfield  {author} {\bibinfo {author} {\bibfnamefont {R.}~\bibnamefont
  {Ferdous}}, \bibinfo {author} {\bibfnamefont {E.}~\bibnamefont {Kawakami}},
  \bibinfo {author} {\bibfnamefont {P.}~\bibnamefont {Scarlino}}, \bibinfo
  {author} {\bibfnamefont {M.~P.}\ \bibnamefont {Nowak}}, \bibinfo {author}
  {\bibfnamefont {D.~R.}\ \bibnamefont {Ward}}, \bibinfo {author}
  {\bibfnamefont {D.~E.}\ \bibnamefont {Savage}}, \bibinfo {author}
  {\bibfnamefont {M.~G.}\ \bibnamefont {Lagally}}, \bibinfo {author}
  {\bibfnamefont {S.~N.}\ \bibnamefont {Coppersmith}}, \bibinfo {author}
  {\bibfnamefont {M.}~\bibnamefont {Friesen}}, \bibinfo {author} {\bibfnamefont
  {M.~A.}\ \bibnamefont {Eriksson}}, \bibinfo {author} {\bibfnamefont
  {L.~M.~K.}\ \bibnamefont {Vandersypen}},\ and\ \bibinfo {author}
  {\bibfnamefont {R.}~\bibnamefont {Rahman}},\ }\bibfield  {title} {\enquote
  {\bibinfo {title} {Valley dependent anisotropic spin splitting in silicon
  quantum dots},}\ }\href {https://doi.org/10.1038/s41534-018-0075-1}
  {\bibfield  {journal} {\bibinfo  {journal} {npj Quantum Inf.}\ }\textbf
  {\bibinfo {volume} {4}},\ \bibinfo {pages} {1} (\bibinfo {year}
  {2018})}\BibitemShut {NoStop}%
\bibitem [{\citenamefont {Hollmann}\ \emph {et~al.}(2020)\citenamefont
  {Hollmann}, \citenamefont {Struck}, \citenamefont {Langrock}, \citenamefont
  {Schmidbauer}, \citenamefont {Schauer}, \citenamefont {Leonhardt},
  \citenamefont {Sawano}, \citenamefont {Riemann}, \citenamefont {Abrosimov},
  \citenamefont {Bougeard},\ and\ \citenamefont
  {Schreiber}}]{hollmann_large_2020}%
  \BibitemOpen
  \bibfield  {author} {\bibinfo {author} {\bibfnamefont {A.}~\bibnamefont
  {Hollmann}}, \bibinfo {author} {\bibfnamefont {T.}~\bibnamefont {Struck}},
  \bibinfo {author} {\bibfnamefont {V.}~\bibnamefont {Langrock}}, \bibinfo
  {author} {\bibfnamefont {A.}~\bibnamefont {Schmidbauer}}, \bibinfo {author}
  {\bibfnamefont {F.}~\bibnamefont {Schauer}}, \bibinfo {author} {\bibfnamefont
  {T.}~\bibnamefont {Leonhardt}}, \bibinfo {author} {\bibfnamefont
  {K.}~\bibnamefont {Sawano}}, \bibinfo {author} {\bibfnamefont
  {H.}~\bibnamefont {Riemann}}, \bibinfo {author} {\bibfnamefont {N.~V.}\
  \bibnamefont {Abrosimov}}, \bibinfo {author} {\bibfnamefont {D.}~\bibnamefont
  {Bougeard}},\ and\ \bibinfo {author} {\bibfnamefont {L.~R.}\ \bibnamefont
  {Schreiber}},\ }\bibfield  {title} {\enquote {\bibinfo {title} {Large,
  {Tunable} {Valley} {Splitting} and {Single}-{Spin} {Relaxation} {Mechanisms}
  in a $\mathrm{Si}$/$\mathrm{Si}_{x}\mathrm{Ge}_{1\ensuremath{-}x}$ {Quantum}
  {Dot}},}\ }\href {https://doi.org/10.1103/PhysRevApplied.13.034068}
  {\bibfield  {journal} {\bibinfo  {journal} {Phys. Rev. Appl.}\ }\textbf
  {\bibinfo {volume} {13}},\ \bibinfo {pages} {034068} (\bibinfo {year}
  {2020})}\BibitemShut {NoStop}%
\bibitem [{\citenamefont {Chen}\ \emph {et~al.}(2021)\citenamefont {Chen},
  \citenamefont {Raach}, \citenamefont {Pan}, \citenamefont {Kiselev},
  \citenamefont {Acuna}, \citenamefont {Blumoff}, \citenamefont {Brecht},
  \citenamefont {Choi}, \citenamefont {Ha}, \citenamefont {Hulbert},
  \citenamefont {Jura}, \citenamefont {Keating}, \citenamefont {Noah},
  \citenamefont {Sun}, \citenamefont {Thomas}, \citenamefont {Borselli},
  \citenamefont {Jackson}, \citenamefont {Rakher},\ and\ \citenamefont
  {Ross}}]{HRL_DAPS}%
  \BibitemOpen
  \bibfield  {author} {\bibinfo {author} {\bibfnamefont {E.~H.}\ \bibnamefont
  {Chen}}, \bibinfo {author} {\bibfnamefont {K.}~\bibnamefont {Raach}},
  \bibinfo {author} {\bibfnamefont {A.}~\bibnamefont {Pan}}, \bibinfo {author}
  {\bibfnamefont {A.~A.}\ \bibnamefont {Kiselev}}, \bibinfo {author}
  {\bibfnamefont {E.}~\bibnamefont {Acuna}}, \bibinfo {author} {\bibfnamefont
  {J.~Z.}\ \bibnamefont {Blumoff}}, \bibinfo {author} {\bibfnamefont
  {T.}~\bibnamefont {Brecht}}, \bibinfo {author} {\bibfnamefont {M.~D.}\
  \bibnamefont {Choi}}, \bibinfo {author} {\bibfnamefont {W.}~\bibnamefont
  {Ha}}, \bibinfo {author} {\bibfnamefont {D.~R.}\ \bibnamefont {Hulbert}},
  \bibinfo {author} {\bibfnamefont {M.~P.}\ \bibnamefont {Jura}}, \bibinfo
  {author} {\bibfnamefont {T.~E.}\ \bibnamefont {Keating}}, \bibinfo {author}
  {\bibfnamefont {R.}~\bibnamefont {Noah}}, \bibinfo {author} {\bibfnamefont
  {B.}~\bibnamefont {Sun}}, \bibinfo {author} {\bibfnamefont {B.~J.}\
  \bibnamefont {Thomas}}, \bibinfo {author} {\bibfnamefont {M.~G.}\
  \bibnamefont {Borselli}}, \bibinfo {author} {\bibfnamefont {C.}~\bibnamefont
  {Jackson}}, \bibinfo {author} {\bibfnamefont {M.~T.}\ \bibnamefont
  {Rakher}},\ and\ \bibinfo {author} {\bibfnamefont {R.~S.}\ \bibnamefont
  {Ross}},\ }\bibfield  {title} {\enquote {\bibinfo {title} {Detuning axis
  pulsed spectroscopy of valley-orbital states in
  $\mathrm{Si}$/$\mathrm{Si}$-$\mathrm{Ge}$ quantum dots},}\ }\href
  {https://doi.org/10.1103/PhysRevApplied.15.044033} {\bibfield  {journal}
  {\bibinfo  {journal} {Phys. Rev. Applied}\ }\textbf {\bibinfo {volume}
  {15}},\ \bibinfo {pages} {044033} (\bibinfo {year} {2021})}\BibitemShut
  {NoStop}%
\bibitem [{\citenamefont {Borjans}\ \emph {et~al.}(2021)\citenamefont
  {Borjans}, \citenamefont {Zhang}, \citenamefont {Mi}, \citenamefont {Cheng},
  \citenamefont {Yao}, \citenamefont {Jackson}, \citenamefont {Edge},\ and\
  \citenamefont {Petta}}]{PRXQuantum.2.020309}%
  \BibitemOpen
  \bibfield  {author} {\bibinfo {author} {\bibfnamefont {F.}~\bibnamefont
  {Borjans}}, \bibinfo {author} {\bibfnamefont {X.}~\bibnamefont {Zhang}},
  \bibinfo {author} {\bibfnamefont {X.}~\bibnamefont {Mi}}, \bibinfo {author}
  {\bibfnamefont {G.}~\bibnamefont {Cheng}}, \bibinfo {author} {\bibfnamefont
  {N.}~\bibnamefont {Yao}}, \bibinfo {author} {\bibfnamefont {C.}~\bibnamefont
  {Jackson}}, \bibinfo {author} {\bibfnamefont {L.}~\bibnamefont {Edge}},\ and\
  \bibinfo {author} {\bibfnamefont {J.}~\bibnamefont {Petta}},\ }\bibfield
  {title} {\enquote {\bibinfo {title} {Probing the variation of the intervalley
  tunnel coupling in a silicon triple quantum dot},}\ }\href
  {https://doi.org/10.1103/PRXQuantum.2.020309} {\bibfield  {journal} {\bibinfo
   {journal} {PRX Quantum}\ }\textbf {\bibinfo {volume} {2}},\ \bibinfo {pages}
  {020309} (\bibinfo {year} {2021})}\BibitemShut {NoStop}%
\bibitem [{\citenamefont {Penthorn}\ \emph {et~al.}(2019)\citenamefont
  {Penthorn}, \citenamefont {Schoenfield}, \citenamefont {Rooney},
  \citenamefont {Edge},\ and\ \citenamefont {Jiang}}]{penthorn}%
  \BibitemOpen
  \bibfield  {author} {\bibinfo {author} {\bibfnamefont {N.~E.}\ \bibnamefont
  {Penthorn}}, \bibinfo {author} {\bibfnamefont {J.~S.}\ \bibnamefont
  {Schoenfield}}, \bibinfo {author} {\bibfnamefont {J.~D.}\ \bibnamefont
  {Rooney}}, \bibinfo {author} {\bibfnamefont {L.~F.}\ \bibnamefont {Edge}},\
  and\ \bibinfo {author} {\bibfnamefont {H.}~\bibnamefont {Jiang}},\ }\bibfield
   {title} {\enquote {\bibinfo {title} {Two-axis quantum control of a fast
  valley qubit in silicon},}\ }\href
  {https://doi.org/10.1038/s41534-019-0212-5} {\bibfield  {journal} {\bibinfo
  {journal} {npj Quantum Inf.}\ }\textbf {\bibinfo {volume} {5}},\ \bibinfo
  {pages} {1} (\bibinfo {year} {2019})}\BibitemShut {NoStop}%
\bibitem [{\citenamefont {McJunkin}\ \emph {et~al.}(2021)\citenamefont
  {McJunkin}, \citenamefont {MacQuarrie}, \citenamefont {Tom}, \citenamefont
  {Neyens}, \citenamefont {Dodson}, \citenamefont {Thorgrimsson}, \citenamefont
  {Corrigan}, \citenamefont {Ercan}, \citenamefont {Savage}, \citenamefont
  {Lagally}, \citenamefont {Joynt}, \citenamefont {Coppersmith}, \citenamefont
  {Friesen},\ and\ \citenamefont {Eriksson}}]{mcjunkin2021valley}%
  \BibitemOpen
  \bibfield  {author} {\bibinfo {author} {\bibfnamefont {T.}~\bibnamefont
  {McJunkin}}, \bibinfo {author} {\bibfnamefont {E.~R.}\ \bibnamefont
  {MacQuarrie}}, \bibinfo {author} {\bibfnamefont {L.}~\bibnamefont {Tom}},
  \bibinfo {author} {\bibfnamefont {S.~F.}\ \bibnamefont {Neyens}}, \bibinfo
  {author} {\bibfnamefont {J.~P.}\ \bibnamefont {Dodson}}, \bibinfo {author}
  {\bibfnamefont {B.}~\bibnamefont {Thorgrimsson}}, \bibinfo {author}
  {\bibfnamefont {J.}~\bibnamefont {Corrigan}}, \bibinfo {author}
  {\bibfnamefont {H.~E.}\ \bibnamefont {Ercan}}, \bibinfo {author}
  {\bibfnamefont {D.~E.}\ \bibnamefont {Savage}}, \bibinfo {author}
  {\bibfnamefont {M.~G.}\ \bibnamefont {Lagally}}, \bibinfo {author}
  {\bibfnamefont {R.}~\bibnamefont {Joynt}}, \bibinfo {author} {\bibfnamefont
  {S.~N.}\ \bibnamefont {Coppersmith}}, \bibinfo {author} {\bibfnamefont
  {M.}~\bibnamefont {Friesen}},\ and\ \bibinfo {author} {\bibfnamefont {M.~A.}\
  \bibnamefont {Eriksson}},\ }\href@noop {} {\enquote {\bibinfo {title}
  {{Valley splittings in Si/SiGe quantum dots with a germanium spike in the
  silicon well}},}\ } (\bibinfo {year} {2021}),\ \Eprint
  {https://arxiv.org/abs/2104.08232} {arXiv:2104.08232} \BibitemShut {NoStop}%
\bibitem [{\citenamefont {Goswami}\ \emph {et~al.}(2007)\citenamefont
  {Goswami}, \citenamefont {Slinker}, \citenamefont {Friesen}, \citenamefont
  {McGuire}, \citenamefont {Truitt}, \citenamefont {Tahan}, \citenamefont
  {Klein}, \citenamefont {Chu}, \citenamefont {Mooney}, \citenamefont {van~der
  Weide}, \citenamefont {Joynt}, \citenamefont {Coppersmith},\ and\
  \citenamefont {Eriksson}}]{goswami_controllable_2007}%
  \BibitemOpen
  \bibfield  {author} {\bibinfo {author} {\bibfnamefont {S.}~\bibnamefont
  {Goswami}}, \bibinfo {author} {\bibfnamefont {K.~A.}\ \bibnamefont
  {Slinker}}, \bibinfo {author} {\bibfnamefont {M.}~\bibnamefont {Friesen}},
  \bibinfo {author} {\bibfnamefont {L.~M.}\ \bibnamefont {McGuire}}, \bibinfo
  {author} {\bibfnamefont {J.~L.}\ \bibnamefont {Truitt}}, \bibinfo {author}
  {\bibfnamefont {C.}~\bibnamefont {Tahan}}, \bibinfo {author} {\bibfnamefont
  {L.~J.}\ \bibnamefont {Klein}}, \bibinfo {author} {\bibfnamefont {J.~O.}\
  \bibnamefont {Chu}}, \bibinfo {author} {\bibfnamefont {P.~M.}\ \bibnamefont
  {Mooney}}, \bibinfo {author} {\bibfnamefont {D.~W.}\ \bibnamefont {van~der
  Weide}}, \bibinfo {author} {\bibfnamefont {R.}~\bibnamefont {Joynt}},
  \bibinfo {author} {\bibfnamefont {S.~N.}\ \bibnamefont {Coppersmith}},\ and\
  \bibinfo {author} {\bibfnamefont {M.~A.}\ \bibnamefont {Eriksson}},\
  }\bibfield  {title} {\enquote {\bibinfo {title} {Controllable valley
  splitting in silicon quantum devices},}\ }\href
  {https://doi.org/10.1038/nphys475} {\bibfield  {journal} {\bibinfo  {journal}
  {Nat. Phys.}\ }\textbf {\bibinfo {volume} {3}},\ \bibinfo {pages} {41}
  (\bibinfo {year} {2007})}\BibitemShut {NoStop}%
\bibitem [{\citenamefont {Yang}\ \emph {et~al.}(2013)\citenamefont {Yang},
  \citenamefont {Rossi}, \citenamefont {Ruskov}, \citenamefont {Lai},
  \citenamefont {Mohiyaddin}, \citenamefont {Lee}, \citenamefont {Tahan},
  \citenamefont {Klimeck}, \citenamefont {Morello},\ and\ \citenamefont
  {Dzurak}}]{yang2013spin}%
  \BibitemOpen
  \bibfield  {author} {\bibinfo {author} {\bibfnamefont {C.~H.}\ \bibnamefont
  {Yang}}, \bibinfo {author} {\bibfnamefont {A.}~\bibnamefont {Rossi}},
  \bibinfo {author} {\bibfnamefont {R.}~\bibnamefont {Ruskov}}, \bibinfo
  {author} {\bibfnamefont {N.~S.}\ \bibnamefont {Lai}}, \bibinfo {author}
  {\bibfnamefont {F.~A.}\ \bibnamefont {Mohiyaddin}}, \bibinfo {author}
  {\bibfnamefont {S.}~\bibnamefont {Lee}}, \bibinfo {author} {\bibfnamefont
  {C.}~\bibnamefont {Tahan}}, \bibinfo {author} {\bibfnamefont
  {G.}~\bibnamefont {Klimeck}}, \bibinfo {author} {\bibfnamefont
  {A.}~\bibnamefont {Morello}},\ and\ \bibinfo {author} {\bibfnamefont {A.~S.}\
  \bibnamefont {Dzurak}},\ }\bibfield  {title} {\enquote {\bibinfo {title}
  {Spin-valley lifetimes in a silicon quantum dot with tunable valley
  splitting},}\ }\href@noop {} {\bibfield  {journal} {\bibinfo  {journal} {Nat.
  Commun.}\ }\textbf {\bibinfo {volume} {4}},\ \bibinfo {pages} {2069}
  (\bibinfo {year} {2013})}\BibitemShut {NoStop}%
\bibitem [{\citenamefont {Jones}\ \emph {et~al.}(2019)\citenamefont {Jones},
  \citenamefont {Pritchett}, \citenamefont {Chen}, \citenamefont {Keating},
  \citenamefont {Andrews}, \citenamefont {Blumoff}, \citenamefont {De~Lorenzo},
  \citenamefont {Eng}, \citenamefont {Ha}, \citenamefont {Kiselev},
  \citenamefont {Meenehan}, \citenamefont {Merkel}, \citenamefont {Wright},
  \citenamefont {Edge}, \citenamefont {Ross}, \citenamefont {Rakher},
  \citenamefont {Borselli},\ and\ \citenamefont {Hunter}}]{HRL_singleshot}%
  \BibitemOpen
  \bibfield  {author} {\bibinfo {author} {\bibfnamefont {A.}~\bibnamefont
  {Jones}}, \bibinfo {author} {\bibfnamefont {E.}~\bibnamefont {Pritchett}},
  \bibinfo {author} {\bibfnamefont {E.}~\bibnamefont {Chen}}, \bibinfo {author}
  {\bibfnamefont {T.}~\bibnamefont {Keating}}, \bibinfo {author} {\bibfnamefont
  {R.}~\bibnamefont {Andrews}}, \bibinfo {author} {\bibfnamefont
  {J.}~\bibnamefont {Blumoff}}, \bibinfo {author} {\bibfnamefont
  {L.}~\bibnamefont {De~Lorenzo}}, \bibinfo {author} {\bibfnamefont
  {K.}~\bibnamefont {Eng}}, \bibinfo {author} {\bibfnamefont {S.}~\bibnamefont
  {Ha}}, \bibinfo {author} {\bibfnamefont {A.}~\bibnamefont {Kiselev}},
  \bibinfo {author} {\bibfnamefont {S.}~\bibnamefont {Meenehan}}, \bibinfo
  {author} {\bibfnamefont {S.}~\bibnamefont {Merkel}}, \bibinfo {author}
  {\bibfnamefont {J.}~\bibnamefont {Wright}}, \bibinfo {author} {\bibfnamefont
  {L.}~\bibnamefont {Edge}}, \bibinfo {author} {\bibfnamefont {R.}~\bibnamefont
  {Ross}}, \bibinfo {author} {\bibfnamefont {M.}~\bibnamefont {Rakher}},
  \bibinfo {author} {\bibfnamefont {M.}~\bibnamefont {Borselli}},\ and\
  \bibinfo {author} {\bibfnamefont {A.}~\bibnamefont {Hunter}},\ }\bibfield
  {title} {\enquote {\bibinfo {title} {Spin-blockade spectroscopy of
  $\mathrm{Si}$/$\mathrm{Si}$-$\mathrm{Ge}$ quantum dots},}\ }\href
  {https://doi.org/10.1103/PhysRevApplied.12.014026} {\bibfield  {journal}
  {\bibinfo  {journal} {Phys. Rev. Applied}\ }\textbf {\bibinfo {volume}
  {12}},\ \bibinfo {pages} {014026} (\bibinfo {year} {2019})}\BibitemShut
  {NoStop}%
\bibitem [{\citenamefont {Shim}\ \emph {et~al.}(2019)\citenamefont {Shim},
  \citenamefont {Ruskov}, \citenamefont {Hurst},\ and\ \citenamefont
  {Tahan}}]{Tahan}%
  \BibitemOpen
  \bibfield  {author} {\bibinfo {author} {\bibfnamefont {Y.-P.}\ \bibnamefont
  {Shim}}, \bibinfo {author} {\bibfnamefont {R.}~\bibnamefont {Ruskov}},
  \bibinfo {author} {\bibfnamefont {H.~M.}\ \bibnamefont {Hurst}},\ and\
  \bibinfo {author} {\bibfnamefont {C.}~\bibnamefont {Tahan}},\ }\bibfield
  {title} {\enquote {\bibinfo {title} {Induced quantum dot probe for material
  characterization},}\ }\href {https://aip.scitation.org/doi/10.1063/1.5053756}
  {\bibfield  {journal} {\bibinfo  {journal} {Appl. Phys. Lett.}\ }\textbf
  {\bibinfo {volume} {114}},\ \bibinfo {pages} {152105} (\bibinfo {year}
  {2019})}\BibitemShut {NoStop}%
\bibitem [{\citenamefont {Topinka}\ \emph {et~al.}(2001)\citenamefont
  {Topinka}, \citenamefont {LeRoy}, \citenamefont {Westervelt}, \citenamefont
  {Shaw}, \citenamefont {Fleischmann}, \citenamefont {Heller}, \citenamefont
  {Maranowski},\ and\ \citenamefont {Gossard}}]{Topinka2001}%
  \BibitemOpen
  \bibfield  {author} {\bibinfo {author} {\bibfnamefont {M.~A.}\ \bibnamefont
  {Topinka}}, \bibinfo {author} {\bibfnamefont {B.~J.}\ \bibnamefont {LeRoy}},
  \bibinfo {author} {\bibfnamefont {R.~M.}\ \bibnamefont {Westervelt}},
  \bibinfo {author} {\bibfnamefont {S.~E.~J.}\ \bibnamefont {Shaw}}, \bibinfo
  {author} {\bibfnamefont {R.}~\bibnamefont {Fleischmann}}, \bibinfo {author}
  {\bibfnamefont {E.~J.}\ \bibnamefont {Heller}}, \bibinfo {author}
  {\bibfnamefont {K.~D.}\ \bibnamefont {Maranowski}},\ and\ \bibinfo {author}
  {\bibfnamefont {A.~C.}\ \bibnamefont {Gossard}},\ }\bibfield  {title}
  {\enquote {\bibinfo {title} {Coherent branched flow in a two-dimensional
  electron gas},}\ }\href {https://doi.org/10.1038/35065553} {\bibfield
  {journal} {\bibinfo  {journal} {Nature}\ }\textbf {\bibinfo {volume} {410}},\
  \bibinfo {pages} {183} (\bibinfo {year} {2001})}\BibitemShut {NoStop}%
\bibitem [{\citenamefont {Barnard}\ \emph {et~al.}(2017)\citenamefont
  {Barnard}, \citenamefont {Hughes}, \citenamefont {Sharpe}, \citenamefont
  {Watanabe}, \citenamefont {Taniguchi},\ and\ \citenamefont
  {Goldhaber-Gordon}}]{Barnard2017}%
  \BibitemOpen
  \bibfield  {author} {\bibinfo {author} {\bibfnamefont {A.~W.}\ \bibnamefont
  {Barnard}}, \bibinfo {author} {\bibfnamefont {A.}~\bibnamefont {Hughes}},
  \bibinfo {author} {\bibfnamefont {A.~L.}\ \bibnamefont {Sharpe}}, \bibinfo
  {author} {\bibfnamefont {K.}~\bibnamefont {Watanabe}}, \bibinfo {author}
  {\bibfnamefont {T.}~\bibnamefont {Taniguchi}},\ and\ \bibinfo {author}
  {\bibfnamefont {D.}~\bibnamefont {Goldhaber-Gordon}},\ }\bibfield  {title}
  {\enquote {\bibinfo {title} {{Absorptive pinhole collimators for ballistic
  Dirac fermions in graphene}},}\ }\href {https://doi.org/10.1038/ncomms15418}
  {\bibfield  {journal} {\bibinfo  {journal} {Nat. Commun.}\ }\textbf {\bibinfo
  {volume} {8}},\ \bibinfo {pages} {15418} (\bibinfo {year}
  {2017})}\BibitemShut {NoStop}%
\bibitem [{\citenamefont {Martin}\ \emph {et~al.}(2008)\citenamefont {Martin},
  \citenamefont {Akerman}, \citenamefont {Ulbricht}, \citenamefont {Lohmann},
  \citenamefont {Smet}, \citenamefont {von Klitzing},\ and\ \citenamefont
  {Yacoby}}]{Martin2008}%
  \BibitemOpen
  \bibfield  {author} {\bibinfo {author} {\bibfnamefont {J.}~\bibnamefont
  {Martin}}, \bibinfo {author} {\bibfnamefont {N.}~\bibnamefont {Akerman}},
  \bibinfo {author} {\bibfnamefont {G.}~\bibnamefont {Ulbricht}}, \bibinfo
  {author} {\bibfnamefont {T.}~\bibnamefont {Lohmann}}, \bibinfo {author}
  {\bibfnamefont {J.~H.}\ \bibnamefont {Smet}}, \bibinfo {author}
  {\bibfnamefont {K.}~\bibnamefont {von Klitzing}},\ and\ \bibinfo {author}
  {\bibfnamefont {A.}~\bibnamefont {Yacoby}},\ }\bibfield  {title} {\enquote
  {\bibinfo {title} {Observation of electron--hole puddles in graphene using a
  scanning single-electron transistor},}\ }\href
  {https://doi.org/10.1038/nphys781} {\bibfield  {journal} {\bibinfo  {journal}
  {Nature Phys.}\ }\textbf {\bibinfo {volume} {4}},\ \bibinfo {pages} {144}
  (\bibinfo {year} {2008})}\BibitemShut {NoStop}%
\bibitem [{\citenamefont {Bastiaans}\ \emph {et~al.}(2019)\citenamefont
  {Bastiaans}, \citenamefont {Cho}, \citenamefont {Chatzopoulos}, \citenamefont
  {Leeuwenhoek}, \citenamefont {Koks},\ and\ \citenamefont
  {Allan}}]{PhysRevB.100.104506}%
  \BibitemOpen
  \bibfield  {author} {\bibinfo {author} {\bibfnamefont {K.~M.}\ \bibnamefont
  {Bastiaans}}, \bibinfo {author} {\bibfnamefont {D.}~\bibnamefont {Cho}},
  \bibinfo {author} {\bibfnamefont {D.}~\bibnamefont {Chatzopoulos}}, \bibinfo
  {author} {\bibfnamefont {M.}~\bibnamefont {Leeuwenhoek}}, \bibinfo {author}
  {\bibfnamefont {C.}~\bibnamefont {Koks}},\ and\ \bibinfo {author}
  {\bibfnamefont {M.~P.}\ \bibnamefont {Allan}},\ }\bibfield  {title} {\enquote
  {\bibinfo {title} {Imaging doubled shot noise in a josephson scanning
  tunneling microscope},}\ }\href {https://doi.org/10.1103/PhysRevB.100.104506}
  {\bibfield  {journal} {\bibinfo  {journal} {Phys. Rev. B}\ }\textbf {\bibinfo
  {volume} {100}},\ \bibinfo {pages} {104506} (\bibinfo {year}
  {2019})}\BibitemShut {NoStop}%
\bibitem [{\citenamefont {Pothier}\ \emph {et~al.}(1997)\citenamefont
  {Pothier}, \citenamefont {Gu\'eron}, \citenamefont {Birge}, \citenamefont
  {Esteve},\ and\ \citenamefont {Devoret}}]{PhysRevLett.79.3490}%
  \BibitemOpen
  \bibfield  {author} {\bibinfo {author} {\bibfnamefont {H.}~\bibnamefont
  {Pothier}}, \bibinfo {author} {\bibfnamefont {S.}~\bibnamefont {Gu\'eron}},
  \bibinfo {author} {\bibfnamefont {N.~O.}\ \bibnamefont {Birge}}, \bibinfo
  {author} {\bibfnamefont {D.}~\bibnamefont {Esteve}},\ and\ \bibinfo {author}
  {\bibfnamefont {M.~H.}\ \bibnamefont {Devoret}},\ }\bibfield  {title}
  {\enquote {\bibinfo {title} {Energy distribution function of quasiparticles
  in mesoscopic wires},}\ }\href {https://doi.org/10.1103/PhysRevLett.79.3490}
  {\bibfield  {journal} {\bibinfo  {journal} {Phys. Rev. Lett.}\ }\textbf
  {\bibinfo {volume} {79}},\ \bibinfo {pages} {3490} (\bibinfo {year}
  {1997})}\BibitemShut {NoStop}%
\bibitem [{\citenamefont {Tikhonov}\ \emph {et~al.}(2020)\citenamefont
  {Tikhonov}, \citenamefont {Denisov}, \citenamefont {Piatrusha}, \citenamefont
  {Khrapach}, \citenamefont {Pekola}, \citenamefont {Karimi}, \citenamefont
  {Jabdaraghi},\ and\ \citenamefont {Khrapai}}]{PhysRevB.102.085417}%
  \BibitemOpen
  \bibfield  {author} {\bibinfo {author} {\bibfnamefont {E.~S.}\ \bibnamefont
  {Tikhonov}}, \bibinfo {author} {\bibfnamefont {A.~O.}\ \bibnamefont
  {Denisov}}, \bibinfo {author} {\bibfnamefont {S.~U.}\ \bibnamefont
  {Piatrusha}}, \bibinfo {author} {\bibfnamefont {I.~N.}\ \bibnamefont
  {Khrapach}}, \bibinfo {author} {\bibfnamefont {J.~P.}\ \bibnamefont
  {Pekola}}, \bibinfo {author} {\bibfnamefont {B.}~\bibnamefont {Karimi}},
  \bibinfo {author} {\bibfnamefont {R.~N.}\ \bibnamefont {Jabdaraghi}},\ and\
  \bibinfo {author} {\bibfnamefont {V.~S.}\ \bibnamefont {Khrapai}},\
  }\bibfield  {title} {\enquote {\bibinfo {title} {Spatial and energy
  resolution of electronic states by shot noise},}\ }\href
  {https://doi.org/10.1103/PhysRevB.102.085417} {\bibfield  {journal} {\bibinfo
   {journal} {Phys. Rev. B}\ }\textbf {\bibinfo {volume} {102}},\ \bibinfo
  {pages} {085417} (\bibinfo {year} {2020})}\BibitemShut {NoStop}%
\bibitem [{\citenamefont {Marguerite}\ \emph {et~al.}(2019)\citenamefont
  {Marguerite}, \citenamefont {Birkbeck}, \citenamefont {Aharon-Steinberg},
  \citenamefont {Halbertal}, \citenamefont {Bagani}, \citenamefont {Marcus},
  \citenamefont {Myasoedov}, \citenamefont {Geim}, \citenamefont {Perello},\
  and\ \citenamefont {Zeldov}}]{Marguerite2019}%
  \BibitemOpen
  \bibfield  {author} {\bibinfo {author} {\bibfnamefont {A.}~\bibnamefont
  {Marguerite}}, \bibinfo {author} {\bibfnamefont {J.}~\bibnamefont
  {Birkbeck}}, \bibinfo {author} {\bibfnamefont {A.}~\bibnamefont
  {Aharon-Steinberg}}, \bibinfo {author} {\bibfnamefont {D.}~\bibnamefont
  {Halbertal}}, \bibinfo {author} {\bibfnamefont {K.}~\bibnamefont {Bagani}},
  \bibinfo {author} {\bibfnamefont {I.}~\bibnamefont {Marcus}}, \bibinfo
  {author} {\bibfnamefont {Y.}~\bibnamefont {Myasoedov}}, \bibinfo {author}
  {\bibfnamefont {A.~K.}\ \bibnamefont {Geim}}, \bibinfo {author}
  {\bibfnamefont {D.~J.}\ \bibnamefont {Perello}},\ and\ \bibinfo {author}
  {\bibfnamefont {E.}~\bibnamefont {Zeldov}},\ }\bibfield  {title} {\enquote
  {\bibinfo {title} {{Imaging work and dissipation in the quantum Hall state in
  graphene}},}\ }\href {https://doi.org/10.1038/s41586-019-1704-3} {\bibfield
  {journal} {\bibinfo  {journal} {Nature}\ }\textbf {\bibinfo {volume} {575}},\
  \bibinfo {pages} {628} (\bibinfo {year} {2019})}\BibitemShut {NoStop}%
\bibitem [{\citenamefont {Uri}\ \emph {et~al.}(2020)\citenamefont {Uri},
  \citenamefont {Grover}, \citenamefont {Cao}, \citenamefont {Crosse},
  \citenamefont {Bagani}, \citenamefont {Rodan-Legrain}, \citenamefont
  {Myasoedov}, \citenamefont {Watanabe}, \citenamefont {Taniguchi},
  \citenamefont {Moon}, \citenamefont {Koshino}, \citenamefont
  {Jarillo-Herrero},\ and\ \citenamefont {Zeldov}}]{Uri2020}%
  \BibitemOpen
  \bibfield  {author} {\bibinfo {author} {\bibfnamefont {A.}~\bibnamefont
  {Uri}}, \bibinfo {author} {\bibfnamefont {S.}~\bibnamefont {Grover}},
  \bibinfo {author} {\bibfnamefont {Y.}~\bibnamefont {Cao}}, \bibinfo {author}
  {\bibfnamefont {J.~A.}\ \bibnamefont {Crosse}}, \bibinfo {author}
  {\bibfnamefont {K.}~\bibnamefont {Bagani}}, \bibinfo {author} {\bibfnamefont
  {D.}~\bibnamefont {Rodan-Legrain}}, \bibinfo {author} {\bibfnamefont
  {Y.}~\bibnamefont {Myasoedov}}, \bibinfo {author} {\bibfnamefont
  {K.}~\bibnamefont {Watanabe}}, \bibinfo {author} {\bibfnamefont
  {T.}~\bibnamefont {Taniguchi}}, \bibinfo {author} {\bibfnamefont
  {P.}~\bibnamefont {Moon}}, \bibinfo {author} {\bibfnamefont {M.}~\bibnamefont
  {Koshino}}, \bibinfo {author} {\bibfnamefont {P.}~\bibnamefont
  {Jarillo-Herrero}},\ and\ \bibinfo {author} {\bibfnamefont {E.}~\bibnamefont
  {Zeldov}},\ }\bibfield  {title} {\enquote {\bibinfo {title} {{Mapping the
  twist-angle disorder and Landau levels in magic-angle graphene}},}\ }\href
  {https://doi.org/10.1038/s41586-020-2255-3} {\bibfield  {journal} {\bibinfo
  {journal} {Nature}\ }\textbf {\bibinfo {volume} {581}},\ \bibinfo {pages}
  {47} (\bibinfo {year} {2020})}\BibitemShut {NoStop}%
\bibitem [{\citenamefont {Ku}\ \emph {et~al.}(2020)\citenamefont {Ku},
  \citenamefont {Zhou}, \citenamefont {Li}, \citenamefont {Shin}, \citenamefont
  {Shi}, \citenamefont {Burch}, \citenamefont {Anderson}, \citenamefont
  {Pierce}, \citenamefont {Xie}, \citenamefont {Hamo}, \citenamefont {Vool},
  \citenamefont {Zhang}, \citenamefont {Casola}, \citenamefont {Taniguchi},
  \citenamefont {Watanabe}, \citenamefont {Fogler}, \citenamefont {Kim},
  \citenamefont {Yacoby},\ and\ \citenamefont {Walsworth}}]{Ku2020}%
  \BibitemOpen
  \bibfield  {author} {\bibinfo {author} {\bibfnamefont {M.~J.~H.}\
  \bibnamefont {Ku}}, \bibinfo {author} {\bibfnamefont {T.~X.}\ \bibnamefont
  {Zhou}}, \bibinfo {author} {\bibfnamefont {Q.}~\bibnamefont {Li}}, \bibinfo
  {author} {\bibfnamefont {Y.~J.}\ \bibnamefont {Shin}}, \bibinfo {author}
  {\bibfnamefont {J.~K.}\ \bibnamefont {Shi}}, \bibinfo {author} {\bibfnamefont
  {C.}~\bibnamefont {Burch}}, \bibinfo {author} {\bibfnamefont {L.~E.}\
  \bibnamefont {Anderson}}, \bibinfo {author} {\bibfnamefont {A.~T.}\
  \bibnamefont {Pierce}}, \bibinfo {author} {\bibfnamefont {Y.}~\bibnamefont
  {Xie}}, \bibinfo {author} {\bibfnamefont {A.}~\bibnamefont {Hamo}}, \bibinfo
  {author} {\bibfnamefont {U.}~\bibnamefont {Vool}}, \bibinfo {author}
  {\bibfnamefont {H.}~\bibnamefont {Zhang}}, \bibinfo {author} {\bibfnamefont
  {F.}~\bibnamefont {Casola}}, \bibinfo {author} {\bibfnamefont
  {T.}~\bibnamefont {Taniguchi}}, \bibinfo {author} {\bibfnamefont
  {K.}~\bibnamefont {Watanabe}}, \bibinfo {author} {\bibfnamefont {M.~M.}\
  \bibnamefont {Fogler}}, \bibinfo {author} {\bibfnamefont {P.}~\bibnamefont
  {Kim}}, \bibinfo {author} {\bibfnamefont {A.}~\bibnamefont {Yacoby}},\ and\
  \bibinfo {author} {\bibfnamefont {R.~L.}\ \bibnamefont {Walsworth}},\
  }\bibfield  {title} {\enquote {\bibinfo {title} {{Imaging viscous flow of the
  Dirac fluid in graphene}},}\ }\href
  {https://doi.org/10.1038/s41586-020-2507-2} {\bibfield  {journal} {\bibinfo
  {journal} {Nature}\ }\textbf {\bibinfo {volume} {583}},\ \bibinfo {pages}
  {537} (\bibinfo {year} {2020})}\BibitemShut {NoStop}%
\bibitem [{\citenamefont {Nowack}\ \emph {et~al.}(2013)\citenamefont {Nowack},
  \citenamefont {Spanton}, \citenamefont {Baenninger}, \citenamefont
  {K{\"o}nig}, \citenamefont {Kirtley}, \citenamefont {Kalisky}, \citenamefont
  {Ames}, \citenamefont {Leubner}, \citenamefont {Br{\"u}ne}, \citenamefont
  {Buhmann}, \citenamefont {Molenkamp}, \citenamefont {Goldhaber-Gordon},\ and\
  \citenamefont {Moler}}]{Nowack2013}%
  \BibitemOpen
  \bibfield  {author} {\bibinfo {author} {\bibfnamefont {K.~C.}\ \bibnamefont
  {Nowack}}, \bibinfo {author} {\bibfnamefont {E.~M.}\ \bibnamefont {Spanton}},
  \bibinfo {author} {\bibfnamefont {M.}~\bibnamefont {Baenninger}}, \bibinfo
  {author} {\bibfnamefont {M.}~\bibnamefont {K{\"o}nig}}, \bibinfo {author}
  {\bibfnamefont {J.~R.}\ \bibnamefont {Kirtley}}, \bibinfo {author}
  {\bibfnamefont {B.}~\bibnamefont {Kalisky}}, \bibinfo {author} {\bibfnamefont
  {C.}~\bibnamefont {Ames}}, \bibinfo {author} {\bibfnamefont {P.}~\bibnamefont
  {Leubner}}, \bibinfo {author} {\bibfnamefont {C.}~\bibnamefont {Br{\"u}ne}},
  \bibinfo {author} {\bibfnamefont {H.}~\bibnamefont {Buhmann}}, \bibinfo
  {author} {\bibfnamefont {L.~W.}\ \bibnamefont {Molenkamp}}, \bibinfo {author}
  {\bibfnamefont {D.}~\bibnamefont {Goldhaber-Gordon}},\ and\ \bibinfo {author}
  {\bibfnamefont {K.~A.}\ \bibnamefont {Moler}},\ }\bibfield  {title} {\enquote
  {\bibinfo {title} {{Imaging currents in HgTe quantum wells in the quantum
  spin Hall regime}},}\ }\href {https://doi.org/10.1038/nmat3682} {\bibfield
  {journal} {\bibinfo  {journal} {Nature Matl.}\ }\textbf {\bibinfo {volume}
  {12}},\ \bibinfo {pages} {787} (\bibinfo {year} {2013})}\BibitemShut
  {NoStop}%
\bibitem [{\citenamefont {Pan}, \citenamefont {Hudson},\ and\ \citenamefont
  {Davis}(1999)}]{Pan_1999}%
  \BibitemOpen
  \bibfield  {author} {\bibinfo {author} {\bibfnamefont {S.~H.}\ \bibnamefont
  {Pan}}, \bibinfo {author} {\bibfnamefont {E.~W.}\ \bibnamefont {Hudson}},\
  and\ \bibinfo {author} {\bibfnamefont {J.~C.}\ \bibnamefont {Davis}},\
  }\bibfield  {title} {\enquote {\bibinfo {title} {{3{H}e refrigerator based
  very low temperature scanning tunneling microscope}},}\ }\href
  {https://aip.scitation.org/doi/10.1063/1.1149605} {\bibfield  {journal}
  {\bibinfo  {journal} {Rev. Sci. Instrum.}\ }\textbf {\bibinfo {volume}
  {70}},\ \bibinfo {pages} {1459} (\bibinfo {year} {1999})}\BibitemShut
  {NoStop}%
\bibitem [{\citenamefont {Song}\ \emph {et~al.}(2010)\citenamefont {Song},
  \citenamefont {Otte}, \citenamefont {Shvarts}, \citenamefont {Zhao},
  \citenamefont {Kuk}, \citenamefont {Blankenship}, \citenamefont {Band},
  \citenamefont {Hess},\ and\ \citenamefont
  {Stroscio}}]{doi:10.1063/1.3520482}%
  \BibitemOpen
  \bibfield  {author} {\bibinfo {author} {\bibfnamefont {Y.~J.}\ \bibnamefont
  {Song}}, \bibinfo {author} {\bibfnamefont {A.~F.}\ \bibnamefont {Otte}},
  \bibinfo {author} {\bibfnamefont {V.}~\bibnamefont {Shvarts}}, \bibinfo
  {author} {\bibfnamefont {Z.}~\bibnamefont {Zhao}}, \bibinfo {author}
  {\bibfnamefont {Y.}~\bibnamefont {Kuk}}, \bibinfo {author} {\bibfnamefont
  {S.~R.}\ \bibnamefont {Blankenship}}, \bibinfo {author} {\bibfnamefont
  {A.}~\bibnamefont {Band}}, \bibinfo {author} {\bibfnamefont {F.~M.}\
  \bibnamefont {Hess}},\ and\ \bibinfo {author} {\bibfnamefont {J.~A.}\
  \bibnamefont {Stroscio}},\ }\bibfield  {title} {\enquote {\bibinfo {title}
  {Invited review article: A 10 mk scanning probe microscopy facility},}\
  }\href@noop {} {\bibfield  {journal} {\bibinfo  {journal} {Rev. Sci.
  Instrum.}\ }\textbf {\bibinfo {volume} {81}},\ \bibinfo {pages} {121101}
  (\bibinfo {year} {2010})}\BibitemShut {NoStop}%
\bibitem [{\citenamefont {Pelliccione}\ \emph {et~al.}(2013)\citenamefont
  {Pelliccione}, \citenamefont {Sciambi}, \citenamefont {Bartel}, \citenamefont
  {Keller},\ and\ \citenamefont {Goldhaber-Gordon}}]{Pelliccione_2013}%
  \BibitemOpen
  \bibfield  {author} {\bibinfo {author} {\bibfnamefont {M.}~\bibnamefont
  {Pelliccione}}, \bibinfo {author} {\bibfnamefont {A.}~\bibnamefont
  {Sciambi}}, \bibinfo {author} {\bibfnamefont {J.}~\bibnamefont {Bartel}},
  \bibinfo {author} {\bibfnamefont {A.}~\bibnamefont {Keller}},\ and\ \bibinfo
  {author} {\bibfnamefont {D.}~\bibnamefont {Goldhaber-Gordon}},\ }\bibfield
  {title} {\enquote {\bibinfo {title} {Design of a scanning gate microscope for
  mesoscopic electron systems in a cryogen-free dilution refrigerator},}\
  }\href {https://aip.scitation.org/doi/citedby/10.1063/1.4794767} {\bibfield
  {journal} {\bibinfo  {journal} {Rev. Sci. Instrum.}\ }\textbf {\bibinfo
  {volume} {84}},\ \bibinfo {pages} {033703} (\bibinfo {year}
  {2013})}\BibitemShut {NoStop}%
\bibitem [{\citenamefont {Hug}\ \emph {et~al.}(1999)\citenamefont {Hug},
  \citenamefont {Stiefel}, \citenamefont {van Schendel}, \citenamefont
  {A.~Moser},\ and\ \citenamefont {Güntherodt}}]{Hug_1999}%
  \BibitemOpen
  \bibfield  {author} {\bibinfo {author} {\bibfnamefont {H.~J.}\ \bibnamefont
  {Hug}}, \bibinfo {author} {\bibfnamefont {B.}~\bibnamefont {Stiefel}},
  \bibinfo {author} {\bibfnamefont {P.~J.~A.}\ \bibnamefont {van Schendel}},
  \bibinfo {author} {\bibfnamefont {S.~M.}\ \bibnamefont {A.~Moser}},\ and\
  \bibinfo {author} {\bibfnamefont {H.-J.}\ \bibnamefont {Güntherodt}},\
  }\bibfield  {title} {\enquote {\bibinfo {title} {A low temperature ultrahigh
  vacuum scanning force microscope},}\ }\href
  {https://aip.scitation.org/doi/10.1063/1.1149970} {\bibfield  {journal}
  {\bibinfo  {journal} {Rev. Sci. Instrum.}\ }\textbf {\bibinfo {volume}
  {70}},\ \bibinfo {pages} {3625} (\bibinfo {year} {1999})}\BibitemShut
  {NoStop}%
\bibitem [{EBL()}]{EBL}%
  \BibitemOpen
  \href@noop {} {}\bibinfo {howpublished}
  {\url{https://eblproducts.com/assets/doc/piezoceramic-tubes.pdf}}\BibitemShut
  {NoStop}%
\bibitem [{\citenamefont {Giessibl}(2019)}]{Giessibl}%
  \BibitemOpen
  \bibfield  {author} {\bibinfo {author} {\bibfnamefont {F.~J.}\ \bibnamefont
  {Giessibl}},\ }\bibfield  {title} {\enquote {\bibinfo {title} {The q{P}lus
  sensor, a powerful core for the atomic force microscope},}\ }\href
  {https://aip.scitation.org/doi/full/10.1063/1.5052264} {\bibfield  {journal}
  {\bibinfo  {journal} {Rev. Sci. Instrum.}\ }\textbf {\bibinfo {volume}
  {90}},\ \bibinfo {pages} {011101} (\bibinfo {year} {2019})}\BibitemShut
  {NoStop}%
\bibitem [{\citenamefont {Castellanos-Gomez}, \citenamefont {Agraït},\ and\
  \citenamefont {Rubio-Bollinger}(2011)}]{Cast}%
  \BibitemOpen
  \bibfield  {author} {\bibinfo {author} {\bibfnamefont {A.}~\bibnamefont
  {Castellanos-Gomez}}, \bibinfo {author} {\bibfnamefont {N.}~\bibnamefont
  {Agraït}},\ and\ \bibinfo {author} {\bibfnamefont {G.}~\bibnamefont
  {Rubio-Bollinger}},\ }\bibfield  {title} {\enquote {\bibinfo {title}
  {Force-gradient-induced mechanical dissipation of quartz tuning fork force
  sensors used in atomic force microscopy},}\ }\href
  {https://doi.org/https://doi.org/10.1016/j.ultramic.2010.11.032} {\bibfield
  {journal} {\bibinfo  {journal} {Ultramicroscopy}\ }\textbf {\bibinfo {volume}
  {111}},\ \bibinfo {pages} {186} (\bibinfo {year} {2011})}\BibitemShut
  {NoStop}%
\bibitem [{\citenamefont {Oliva}\ \emph {et~al.}(1996)\citenamefont {Oliva},
  \citenamefont {Romero~G.}, \citenamefont {Peña}, \citenamefont {Anguiano},\
  and\ \citenamefont {Aguilar}}]{oliva}%
  \BibitemOpen
  \bibfield  {author} {\bibinfo {author} {\bibfnamefont {A.~I.}\ \bibnamefont
  {Oliva}}, \bibinfo {author} {\bibfnamefont {A.}~\bibnamefont {Romero~G.}},
  \bibinfo {author} {\bibfnamefont {J.~L.}\ \bibnamefont {Peña}}, \bibinfo
  {author} {\bibfnamefont {E.}~\bibnamefont {Anguiano}},\ and\ \bibinfo
  {author} {\bibfnamefont {M.}~\bibnamefont {Aguilar}},\ }\bibfield  {title}
  {\enquote {\bibinfo {title} {Electrochemical preparation of tungsten tips for
  a scanning tunneling microscope},}\ }\href@noop {} {\bibfield  {journal}
  {\bibinfo  {journal} {Rev. Sci. Instrum.}\ }\textbf {\bibinfo {volume}
  {67}},\ \bibinfo {pages} {1917} (\bibinfo {year} {1996})}\BibitemShut
  {NoStop}%
\bibitem [{\citenamefont {Shperber}\ \emph {et~al.}(2019)\citenamefont
  {Shperber}, \citenamefont {Vardi}, \citenamefont {Persky}, \citenamefont
  {Wissberg}, \citenamefont {Huber},\ and\ \citenamefont {Kalisky}}]{Kalisky}%
  \BibitemOpen
  \bibfield  {author} {\bibinfo {author} {\bibfnamefont {Y.}~\bibnamefont
  {Shperber}}, \bibinfo {author} {\bibfnamefont {N.}~\bibnamefont {Vardi}},
  \bibinfo {author} {\bibfnamefont {E.}~\bibnamefont {Persky}}, \bibinfo
  {author} {\bibfnamefont {S.}~\bibnamefont {Wissberg}}, \bibinfo {author}
  {\bibfnamefont {M.~E.}\ \bibnamefont {Huber}},\ and\ \bibinfo {author}
  {\bibfnamefont {B.}~\bibnamefont {Kalisky}},\ }\bibfield  {title} {\enquote
  {\bibinfo {title} {Scanning {SQUID} microscopy in a cryogen-free cooler},}\
  }\href {https://aip.scitation.org/doi/full/10.1063/1.5087060} {\bibfield
  {journal} {\bibinfo  {journal} {Rev. Sci. Instrum.}\ }\textbf {\bibinfo
  {volume} {90}},\ \bibinfo {pages} {053702} (\bibinfo {year}
  {2019})}\BibitemShut {NoStop}%
\bibitem [{\citenamefont {Geaney}\ \emph {et~al.}(2019)\citenamefont {Geaney},
  \citenamefont {Cox}, \citenamefont {H{\"o}nigl-Decrinis}, \citenamefont
  {Shaikhaidarov}, \citenamefont {Kubatkin}, \citenamefont {Lindstr{\"o}m},
  \citenamefont {Danilov},\ and\ \citenamefont {de~Graaf}}]{Geaney2019}%
  \BibitemOpen
  \bibfield  {author} {\bibinfo {author} {\bibfnamefont {S.}~\bibnamefont
  {Geaney}}, \bibinfo {author} {\bibfnamefont {D.}~\bibnamefont {Cox}},
  \bibinfo {author} {\bibfnamefont {T.}~\bibnamefont {H{\"o}nigl-Decrinis}},
  \bibinfo {author} {\bibfnamefont {R.}~\bibnamefont {Shaikhaidarov}}, \bibinfo
  {author} {\bibfnamefont {S.~E.}\ \bibnamefont {Kubatkin}}, \bibinfo {author}
  {\bibfnamefont {T.}~\bibnamefont {Lindstr{\"o}m}}, \bibinfo {author}
  {\bibfnamefont {A.~V.}\ \bibnamefont {Danilov}},\ and\ \bibinfo {author}
  {\bibfnamefont {S.~E.}\ \bibnamefont {de~Graaf}},\ }\bibfield  {title}
  {\enquote {\bibinfo {title} {Near-field scanning microwave microscopy in the
  single photon regime},}\ }\href {https://doi.org/10.1038/s41598-019-48780-3}
  {\bibfield  {journal} {\bibinfo  {journal} {Sci. Rep.}\ }\textbf {\bibinfo
  {volume} {9}},\ \bibinfo {pages} {12539} (\bibinfo {year}
  {2019})}\BibitemShut {NoStop}%
\bibitem [{\citenamefont {Woodside}\ and\ \citenamefont
  {McEuen}(2002)}]{Woodside_2002}%
  \BibitemOpen
  \bibfield  {author} {\bibinfo {author} {\bibfnamefont {M.~T.}\ \bibnamefont
  {Woodside}}\ and\ \bibinfo {author} {\bibfnamefont {P.~L.}\ \bibnamefont
  {McEuen}},\ }\bibfield  {title} {\enquote {\bibinfo {title} {Scanned probe
  imaging of single-electron charge states in nanotube quantum dots},}\ }\href
  {https://doi.org/10.1126/science.1069923} {\bibfield  {journal} {\bibinfo
  {journal} {Science}\ }\textbf {\bibinfo {volume} {296}},\ \bibinfo {pages}
  {1098} (\bibinfo {year} {2002})}\BibitemShut {NoStop}%
\bibitem [{\citenamefont {Pioda}\ \emph {et~al.}(2004)\citenamefont {Pioda},
  \citenamefont {Ki\ifmmode~\check{c}\else \v{c}\fi{}in}, \citenamefont {Ihn},
  \citenamefont {Sigrist}, \citenamefont {Fuhrer}, \citenamefont {Ensslin},
  \citenamefont {Weichselbaum}, \citenamefont {Ulloa}, \citenamefont
  {Reinwald},\ and\ \citenamefont {Wegscheider}}]{PhysRevLett.93.216801}%
  \BibitemOpen
  \bibfield  {author} {\bibinfo {author} {\bibfnamefont {A.}~\bibnamefont
  {Pioda}}, \bibinfo {author} {\bibfnamefont {S.}~\bibnamefont
  {Ki\ifmmode~\check{c}\else \v{c}\fi{}in}}, \bibinfo {author} {\bibfnamefont
  {T.}~\bibnamefont {Ihn}}, \bibinfo {author} {\bibfnamefont {M.}~\bibnamefont
  {Sigrist}}, \bibinfo {author} {\bibfnamefont {A.}~\bibnamefont {Fuhrer}},
  \bibinfo {author} {\bibfnamefont {K.}~\bibnamefont {Ensslin}}, \bibinfo
  {author} {\bibfnamefont {A.}~\bibnamefont {Weichselbaum}}, \bibinfo {author}
  {\bibfnamefont {S.~E.}\ \bibnamefont {Ulloa}}, \bibinfo {author}
  {\bibfnamefont {M.}~\bibnamefont {Reinwald}},\ and\ \bibinfo {author}
  {\bibfnamefont {W.}~\bibnamefont {Wegscheider}},\ }\bibfield  {title}
  {\enquote {\bibinfo {title} {Spatially resolved manipulation of single
  electrons in quantum dots using a scanned probe},}\ }\href
  {https://doi.org/10.1103/PhysRevLett.93.216801} {\bibfield  {journal}
  {\bibinfo  {journal} {Phys. Rev. Lett.}\ }\textbf {\bibinfo {volume} {93}},\
  \bibinfo {pages} {216801} (\bibinfo {year} {2004})}\BibitemShut {NoStop}%
\bibitem [{\citenamefont {Crook}\ \emph {et~al.}(2003)\citenamefont {Crook},
  \citenamefont {Smith}, \citenamefont {Graham}, \citenamefont {Farrer},
  \citenamefont {Beere},\ and\ \citenamefont
  {Ritchie}}]{PhysRevLett.91.246803}%
  \BibitemOpen
  \bibfield  {author} {\bibinfo {author} {\bibfnamefont {R.}~\bibnamefont
  {Crook}}, \bibinfo {author} {\bibfnamefont {C.~G.}\ \bibnamefont {Smith}},
  \bibinfo {author} {\bibfnamefont {A.~C.}\ \bibnamefont {Graham}}, \bibinfo
  {author} {\bibfnamefont {I.}~\bibnamefont {Farrer}}, \bibinfo {author}
  {\bibfnamefont {H.~E.}\ \bibnamefont {Beere}},\ and\ \bibinfo {author}
  {\bibfnamefont {D.~A.}\ \bibnamefont {Ritchie}},\ }\bibfield  {title}
  {\enquote {\bibinfo {title} {Imaging fractal conductance fluctuations and
  scarred wave functions in a quantum billiard},}\ }\href
  {https://doi.org/10.1103/PhysRevLett.91.246803} {\bibfield  {journal}
  {\bibinfo  {journal} {Phys. Rev. Lett.}\ }\textbf {\bibinfo {volume} {91}},\
  \bibinfo {pages} {246803} (\bibinfo {year} {2003})}\BibitemShut {NoStop}%
\bibitem [{\citenamefont {Mi}\ \emph {et~al.}(2015)\citenamefont {Mi},
  \citenamefont {Hazard}, \citenamefont {Payette}, \citenamefont {Wang},
  \citenamefont {Zajac}, \citenamefont {Cady},\ and\ \citenamefont
  {Petta}}]{xiao_mobility}%
  \BibitemOpen
  \bibfield  {author} {\bibinfo {author} {\bibfnamefont {X.}~\bibnamefont
  {Mi}}, \bibinfo {author} {\bibfnamefont {T.~M.}\ \bibnamefont {Hazard}},
  \bibinfo {author} {\bibfnamefont {C.}~\bibnamefont {Payette}}, \bibinfo
  {author} {\bibfnamefont {K.}~\bibnamefont {Wang}}, \bibinfo {author}
  {\bibfnamefont {D.~M.}\ \bibnamefont {Zajac}}, \bibinfo {author}
  {\bibfnamefont {J.~V.}\ \bibnamefont {Cady}},\ and\ \bibinfo {author}
  {\bibfnamefont {J.~R.}\ \bibnamefont {Petta}},\ }\bibfield  {title} {\enquote
  {\bibinfo {title} {Magnetotransport studies of mobility limiting mechanisms
  in undoped {Si/SiGe} heterostructures},}\ }\href
  {https://doi.org/10.1103/PhysRevB.92.035304} {\bibfield  {journal} {\bibinfo
  {journal} {Phys. Rev. B}\ }\textbf {\bibinfo {volume} {92}},\ \bibinfo
  {pages} {035304} (\bibinfo {year} {2015})}\BibitemShut {NoStop}%
\bibitem [{\citenamefont {Hanson}\ \emph {et~al.}(2007)\citenamefont {Hanson},
  \citenamefont {Kouwenhoven}, \citenamefont {Petta}, \citenamefont {Tarucha},\
  and\ \citenamefont {Vandersypen}}]{RevModPhys.79.1217}%
  \BibitemOpen
  \bibfield  {author} {\bibinfo {author} {\bibfnamefont {R.}~\bibnamefont
  {Hanson}}, \bibinfo {author} {\bibfnamefont {L.~P.}\ \bibnamefont
  {Kouwenhoven}}, \bibinfo {author} {\bibfnamefont {J.~R.}\ \bibnamefont
  {Petta}}, \bibinfo {author} {\bibfnamefont {S.}~\bibnamefont {Tarucha}},\
  and\ \bibinfo {author} {\bibfnamefont {L.~M.~K.}\ \bibnamefont
  {Vandersypen}},\ }\bibfield  {title} {\enquote {\bibinfo {title} {Spins in
  few-electron quantum dots},}\ }\href
  {https://doi.org/10.1103/RevModPhys.79.1217} {\bibfield  {journal} {\bibinfo
  {journal} {Rev. Mod. Phys.}\ }\textbf {\bibinfo {volume} {79}},\ \bibinfo
  {pages} {1217} (\bibinfo {year} {2007})}\BibitemShut {NoStop}%
\bibitem [{\citenamefont {Fallahi}\ \emph {et~al.}(2005)\citenamefont
  {Fallahi}, \citenamefont {Bleszynski}, \citenamefont {Westervelt},
  \citenamefont {Huang}, \citenamefont {Walls}, \citenamefont {Heller},
  \citenamefont {Hanson},\ and\ \citenamefont {Gossard}}]{Fallahi2005}%
  \BibitemOpen
  \bibfield  {author} {\bibinfo {author} {\bibfnamefont {P.}~\bibnamefont
  {Fallahi}}, \bibinfo {author} {\bibfnamefont {A.~C.}\ \bibnamefont
  {Bleszynski}}, \bibinfo {author} {\bibfnamefont {R.~M.}\ \bibnamefont
  {Westervelt}}, \bibinfo {author} {\bibfnamefont {J.}~\bibnamefont {Huang}},
  \bibinfo {author} {\bibfnamefont {J.~D.}\ \bibnamefont {Walls}}, \bibinfo
  {author} {\bibfnamefont {E.~J.}\ \bibnamefont {Heller}}, \bibinfo {author}
  {\bibfnamefont {M.}~\bibnamefont {Hanson}},\ and\ \bibinfo {author}
  {\bibfnamefont {A.~C.}\ \bibnamefont {Gossard}},\ }\bibfield  {title}
  {\enquote {\bibinfo {title} {Imaging a single-electron quantum dot},}\ }\href
  {https://doi.org/10.1021/nl048405v} {\bibfield  {journal} {\bibinfo
  {journal} {Nano Lett.}\ }\textbf {\bibinfo {volume} {5}},\ \bibinfo {pages}
  {223} (\bibinfo {year} {2005})}\BibitemShut {NoStop}%
\end{thebibliography}%
\end{document}